%% file: main.tex
\pdfoutput=1 
\documentclass[11pt,a4paper]{article}

\usepackage{jheppub} 

\usepackage[T1]{fontenc} 
 
\usepackage{amsmath,amsthm,amssymb}
\usepackage{multicol}
\usepackage{blkarray}

\newcommand{\Z}{\mathbb{Z}}

\newcommand{\C}{\mathbb{C}}

\newcommand{\imrho}{\text{Im}\rho}

\newcommand{\V}{\mathcal{V}}
\newcommand{\antiD}{\overline{\text{D3}}}

\numberwithin{equation}{section}

\newcommand*{\id}{{\normalfont\hbox{1\kern-0.15em \vrule width .8pt depth-.5pt}}}

\usepackage[utf8]{inputenc}
\usepackage[english]{babel}
\usepackage{textcomp}
\usepackage{amsmath}
\usepackage{mathtools}
\usepackage{gensymb}
\usepackage{multirow}
\usepackage{multicol}
\usepackage{amsbsy}
\usepackage{amssymb}
\usepackage{capt-of}
\usepackage[T1]{fontenc} 
\usepackage{physics}
\usepackage{multicol} 
\usepackage{multirow}
\usepackage[table]{xcolor}
\usepackage{lscape}
\usepackage[hang, small,labelfont=bf,up]{caption} \usepackage[title]{appendix}
\usepackage[format=plain,
            labelfont={bf}]{caption}

\def\be{\begin{equation}}
\def\ee{\end{equation}}
\def\bea{\begin{eqnarray}}
\def\eea{\end{eqnarray}}
\def\nn{\nonumber}
\def\ni{\noindent}

\usepackage{booktabs} 
\usepackage{float} 

\usepackage{hyperref} 
\usepackage{graphicx} 
\usepackage{epstopdf} 
\usepackage{paralist} 
\usepackage{subcaption}
\usepackage[framemethod=TikZ]{mdframed}

\mdfdefinestyle{Calculation}{%
    linecolor=black!80!white,
    outerlinewidth=0.5pt,
    roundcorner=0pt,
    innertopmargin=15pt,
    innerbottommargin=15pt,
    innerrightmargin=20pt,
    innerleftmargin=20pt,
    backgroundcolor=yellow!15!white}

\mdfdefinestyle{Example}{%
    linecolor=black!80!white,
    outerlinewidth=0.5pt,
    roundcorner=0pt,
    innertopmargin=15pt,
    innerbottommargin=15pt,
    innerrightmargin=20pt,
    innerleftmargin=20pt,
    backgroundcolor=blue!5!white}
    
\RequirePackage{color}
\usepackage{colortbl}
\definecolor{green2}{cmyk}{0, 1, 0.5, 0.3}
\definecolor{green3}{cmyk}{1, 0.75, 1.0, 0.0}

\definecolor{lightgreen}{cmyk}{0.2, 0, 0.2, 0.2}
\definecolor{lightgray}{cmyk}{0.1,0.2,0,0.1}
\definecolor{lightgray2}{cmyk}{0.4,0.4,0,0.8}
\definecolor{black}{cmyk}{1.0,1.0,1.0,1.0} 
  \hypersetup{
    colorlinks=true,
    linkbordercolor={white},
    linkcolor={green3},
    citecolor={green2},
    filecolor={green2},      
    urlcolor={magenta},
}

\title{\huge A New de Sitter Solution with a\\Weakly Warped Deformed Conifold}

\author[a]{Bruno Valeixo Bento,}
\author[b]{Dibya Chakraborty,}
\author[a]{Susha Parameswaran,}
\author[c]{Ivonne Zavala}

\affiliation[a]{Department of Mathematical Sciences, University of Liverpool, Liverpool L69 7ZL}
\affiliation[b]{Departamento de Física, Universidad de Guanajuato, Loma del Bosque No. 103
        Col. Lomas del Campestre, C.P 37150 León, Guanajuato, México}
\affiliation[c]{Physics Department, Swansea University, SA2 8PP, UK}

\emailAdd{Bruno.Bento@liv.ac.uk}
\emailAdd{dibyac@fisica.ugto.mx}
\emailAdd{susha@liv.ac.uk}
\emailAdd{e.i.zavalacarrasco@swansea.ac.uk}

\abstract{We revisit moduli stabilisation for type IIB flux compactifications that include a warped throat region corresponding to a warped deformed conifold, with an anti-D3-brane sitting at its tip.  The warping induces a coupling between the conifold's deformation modulus and the bulk volume modulus in the K\"ahler potential.  Previous works have studied the scalar potential assuming a strong warping such that this coupling term dominates, and found that the anti-D3-brane uplift may destabilise the conifold modulus and/or volume modulus, unless flux numbers within the throat are large, which makes tadpole cancellation a challenge.  We explore the regime of parameter space corresponding to a weakly-but-still warped throat, such that the coupling between the conifold and volume moduli is subdominant.  We thus discover a new metastable de Sitter solution within the four-dimensional effective field theory. We discuss the position of this de Sitter vacuum in the string theory landscape and swampland.}

\begin{document}
\maketitle

\input{Sections/introduction}

\input{Sections/potential}
\input{Sections/DeformationModulusStabilisation}
\input{Sections/LVS}
\input{Sections/conclusions}

\section*{Acknowledgements}
We are grateful to Nana Cabo-Bizet, Chiara Crin\'o, Mariana Gra\~na, Carlos N\'u\~nez, Severin L\"ust, Gianmassimo Tasinato and Flavio Tonioni for discussions. D. C. thanks the Theoretical Physics group at University of Liverpool for great hospitality and support while part of this work was done. IZ is partially supported by STFC, grant ST/P00055X/1. D. C. is supported by a CONACyT Mexico grant, Beca-Mixta CONACyT and DAIP-UG (2018/2019). 

\newpage
\appendix
\input{Sections/cs_moduli_metric}

\bibliographystyle{utphys}

\bibliography{references}

\end{document}

%% file: Sections/introduction.tex
\section{Introduction}
\label{S:Introduction}

Cosmological observations on Dark Energy are thus far consistent with a cosmological constant sourcing a de Sitter Universe \cite{Planck}.  However, almost two decades after the seminal paper by Kachru, Kallosh, Linde and Trivedi (KKLT) \cite{KKLT}, which provided a path towards constructing de Sitter vacua in string theory, it remains an open question as to whether there is a vast landscape of metastable de Sitter vacua in string theory or none at all (see e.g. \cite{Sethi2017phn,Danielsson2018,Garg2018,Cicoli2018,Ooguri2018} and the references therein).  This paper contributes to the ongoing debate on moduli stabilisation to de Sitter vacua in type IIB string theory, using the interplay between fluxes, quantum corrections and an anti-D3-brane uplift.  It studies type IIB flux compactifications that include an anti-D3-brane at the tip of a warped throat region.  Deep in the throat region, the geometry is well-described by the Klebanov-Strassler (KS) solution, known also as the warped deformed conifold.  By considering the effect of gluing the warped throat onto a compact Calabi-Yau, we investigate the stabilisation of the conifold and volume moduli, which are coupled.  

This program was initiated in \cite{upliftingrunaways2019} (see \cite{Blumenhagen:2019qcg, Bena:2019sxm, Dudas:2019pls, lisa2019,Blumenhagen:2020dea,Seo:2021kyi} for further work), where it was noticed that the uplifting potential energy contribution from a probe anti-D3-brane at the tip of the warped throat can destabilise the conifold modulus that corresponds to a deformation of the tip, causing it to runaway to its singular point.  To avoid this runaway and ensure that the potential has a critical point, a minimum flux is required, namely $\sqrt{g_s} M \gtrsim 6.8$, with $M$ the RR-flux on the deformed $S^3$ at the tip.  At the same time, in order that the uplifting brane potential energy be sufficiently suppressed to avoid a runaway in the volume modulus, the warp factor at the tip, $e^{-4A_0} \sim e^{\frac{8\pi K}{3g_s M}}$, must be sufficiently strong, with $K$ the NSNS-flux dual to $M$.  Together, large $\sqrt{g_s} M$ and large $\frac{8\pi K}{g_s M}$ imply large $M K$ and a large positive contribution to the D3-tadpole in the compact internal manifold, which may be difficult to cancel within perturbative type IIB string theory \cite{upliftingrunaways2019}.  The bound on $M K$, and other flux numbers required to stabilise bulk moduli, was further refined in \cite{Braun:2020jrx,tadpoleProblem,Bena:2021tadpole}, with $M K \gtrsim 500$.

The same issue was studied recently for the Large Volume Scenario in \cite{LVSdS:2010.15903}.  Again, a lower bound on the flux numbers was found necessary to avoid the conifold runaway, and moreover, it was emphasised that an even stronger bound on the D3-tadpole contribution emerges from the supergravity approximation, $g_s M \gg 1$.  The nature of the AdS pre-uplift-vacuum in LVS is such that the hierarchy required to maintain volume stabilisation after uplift is somewhat smaller, leading to a rough bound of $M K \gtrsim 100$. Examples of metastable dS vacua were constructed, achieving cancellation of the D3-tadpole with the introduction of so-called Whitney branes \cite{Collinucci2008,Collinucci2008sq}.

In this paper we consider a new, previously unexplored, region of parameter space for the interplay between the volume and conifold moduli.  In particular, in the presence of warping, there is an extra contribution to the K\"ahler metric of the conifold modulus, which includes a dependence on the volume modulus, and it is this mixing that has been studied in \cite{upliftingrunaways2019, Blumenhagen:2019qcg, Bena:2019sxm, Dudas:2019pls, lisa2019, ScalisiConifold}. However, depending on the balance between the strength of warping, the large bulk volume and the small conifold deformation, the warping contribution may  be subdominant in the K\"ahler metric.  By considering this latter weakly-but-still-warped regime, we discover a new metastable de Sitter solution within the four-dimensional effective field theory (EFT).  Like the LVS de Sitter vacuum in \cite{LVSdS:2010.15903}, the throat fluxes are $M K \gtrsim 100$, there is no bulk-singularity problem \cite{Carta_2019, Gao:2020xqh}, and the solution is consistent with the $\alpha'$ and string-loop expansions and the four-dimensional supergravity description.

The paper is organised as follows.  In the next section \ref{sec:uplifting_runaways}, we give a brief review of the volume modulus and the conifold deformation modulus and how they appear in the 4d low energy effective field theory description of type IIB warped flux compactifications with an $\antiD$-brane.  In Section \ref{DeformStab_sec}, we study the stabilisation of the conifold modulus in two regimes, the one where the warping term dominates in its K\"ahler metric and the one where it is subdominant.  In the strongly warped regime, we recover the solution in \cite{upliftingrunaways2019}, and in the weakly warped one we find a new solution.  In Section \ref{LVS_section}, we extend this analysis to include the volume modulus stabilisation, using the Large Volume Scenario in both regimes.  Again, in the regime of strong warping we recover the LVS solution in \cite{LVSdS:2010.15903}, and when the warping is subdominant we find a new metastable de Sitter solution.  We conclude in Section \ref{S:Conclusions} with a brief summary and a discussion on possible control issues for our new de Sitter solution.  In Appendix \ref{sec:complexstructuremoduli} we review the derivation of the conifold modulus's K\"ahler metric, including its dependence on the volume modulus, based on  \cite{giddings2003scales,douglas2007warping}.

%% file: Sections/potential.tex
\section{The Coupled Volume and Conifold Deformation Moduli}
\label{sec:uplifting_runaways}
In this section we review type IIB flux compactifications with warped throats and their 4d effective field theory description, paying special attention to the volume modulus and the complex structure modulus that corresponds to the deformation of a Klebanov-Strassler throat.  In particular we  highlight the identification of the volume modulus and its mixing with the conifold deformation modulus, which occurs in the K\"ahler potential of the 4d EFT, following dimensional reduction.  At the same time, we review the important mass scales.  These results will be used in the following sections to study the stabilisation of the moduli. 

\subsection{The moduli and mass scales}

We start from the type IIB supergravity bosonic action in the Einstein-frame\footnote{In our conventions, the string frame and Einstein frame 10d metrics are related by $G_{MN}^S = e^{\frac{\phi-\phi_0}{2}}G_{MN}^E$, where $e^{\phi_0}=g_s$ with $\phi_0$ being the background value of the dilaton. This means $\kappa = \kappa_{10} g_s$ and hence $2\kappa^2 = (2\pi)^7g_s^2\alpha'^4$. \label{F:vev_shift}}
\begin{align}
    S_{IIB}^{boson} = \frac{1}{2\kappa^2}\int d^{10}x \sqrt{-g_{10}} \left\{R_{10} - \frac{\partial_M\tau\partial^M\Bar{\tau}}{2(\text{Im}\tau)^2} - \frac{g_s|G_3|^2}{2(\text{Im}\tau)} - \frac{g_s^2|F_5|^2}{4} \right\}
    - \frac{ig_s^2}{8\kappa^2}\int \frac{C_4^+\wedge G_3\wedge \Bar{G}_3}{\text{Im}\tau} .
    \label{eq:TypeIIB}
\end{align}
We consider a flux compactification, whose internal compact space consists of a finite portion of a warped throat --- a warped non-singular deformed conifold \cite{KS2000supergravity} --- glued to a compact Calabi-Yau.  The 10d metric can be written as:
\begin{align}
    ds^2 &= e^{2A} e^{2\Omega(x)} g_{\mu\nu}dx^{\mu}dx^{\nu} + e^{-2A} g_{mn}dy^mdy^n,
    \label{eq:metric}
\end{align}
where the factor $e^{2\Omega(x)}$ anticipates the Weyl-rescaling which is necessary to go to the Einstein frame in 4d.  The background warp factor, $e^{-4A}$, that solves the 10d Einstein equations in the presence of fluxes, is only fixed up to a constant shift, motivating the following form \cite{frey2009universal}
\begin{align}
    e^{-4A(y)} = e^{-4A_0(y)} + c.
\end{align}
The fact that $g_{mn}\rightarrow\lambda g_{mn}$ together with $e^{2A}\rightarrow \lambda e^{2A}$ is a gauge redundancy of the metric \cite{Giddings:2005ff,Aparicio:2015psl}, allows us to choose $\lambda=c(x)^{1/2}$ and rewrite (\ref{eq:metric}) as
\begin{align}
    ds^2 
    &= \left[1 + \frac{e^{-4A_0(y)}}{c(x)}\right]^{-1/2} e^{2\Omega(x)} g_{\mu\nu}dx^{\mu}dx^{\nu}  + \left[1 + \frac{e^{-4A_0(y)}}{c(x)}\right]^{1/2}  c(x)^{1/2} g_{mn}dy^mdy^n \,,
    \label{eq:10dmetric}
\end{align}
and therefore define the warp factor as 
\be\label{WF}
h=1+\frac{e^{-4A_0(y)}}{c(x)} \,.
\ee
With this definition, we naturally recover the unwarped case in the $c(x)\rightarrow\infty$ limit, when $h=1$. In that limit we identify $c(x)=\mathcal{V}^{2/3}$, with $\mathcal{V} l_s^6$ the unwarped volume of the compact space, as we would expect for a volume modulus, where we have chosen coordinates such that $V_{6} = \int d^6y\sqrt{g_6} = l_s^6$. 

Dimensionally reducing the 10d Einstein-Hilbert term in (\ref{eq:TypeIIB}) down to 4d using this metric, we obtain
\begin{align}
    S  &= \frac{V_w^0}{2\kappa^2}\int d^{4}x \sqrt{-g_4}  R_4 + ... \,,
    \label{eq:dimRed_kineticTerms}
\end{align}
where we have defined the Weyl factor $e^{2\Omega}$ as 
\begin{align}
    e^{2\Omega(x)} = \frac{V_w^0}{c(x)^{3/2}\int d^6y \sqrt{g_6} ~h} = \frac{V_w^0}{V_w},
    && V_w = ~c(x)^{3/2}\int d^6y \sqrt{g_6} ~h\,.
	\label{eq:Vw_definition}
\end{align}
The choice of $V_w^0$, which ensures that the Weyl factor is dimensionless, is arbitrary\footnote{Note that all the mass-scales in units of $M_{p}$ will be independent of the normalisation of $e^{2\Omega(x)}$.}. However, the most convenient choice is $V_w^0 \equiv \langle V_w \rangle$, such that $\langle e^{2\Omega(x)}\rangle =1$, i.e. the two frames are the same at the vev, which allows us to relate the string scale with the 4d Planck scale using frame independent volumes.  In the limit of small warping (or for no warping at all), where $c(x)\gg e^{-4A_0(y)}$ and $h\approx 1$ for most of the internal space, we have $V_w^0 \approx \langle\mathcal{V}\rangle l_s^6$, the physical volume of the compact space. With our choice of Weyl-rescaling, the definition of $M_p$ is
\begin{align}
    \frac{M_p^2}{2} = \frac{1}{2\kappa_4^2} = \frac{V_w^0}{2\kappa^2}
	\implies
	\frac{m_s}{M_p} = \frac{g_s}{\sqrt{4\pi\V_w^0}} \,,
	\label{eq:msMp}
\end{align}
where $ V_w^0 =\V_w^0 l_s^6$. Note that the relation between the string scale\footnote{In our conventions, $(2\pi)^2\alpha' = l_s^2$ and $m_s=l_s^{-1}$, which differs from $M_s=1/\sqrt{\alpha'} = 2\pi m_s$.} and the 4d Planck scale is only determined once the volume modulus is fixed (analogously to what happens with the dilaton in 10d, where the physical gravitational coupling can only be related to the string frame coupling once the dilaton is fixed).  Therefore, when studying the moduli stabilisation within the 4d EFT, we replace any string scale with the 4d Planck scale, $M_{p}$. 

The mass-squareds associated with the Kaluza-Klein towers of massive states arising from the compactification can be derived from the Laplacian \cite{quevedo2008warped}
\begin{align}
\Delta = -\frac{e^{2\Omega}}{c^{1/2}}\frac{h^{1/2}}{\sqrt{g}} \partial_m\left(\sqrt{g}h^{-1}g^{mn} \partial_n \right) \,, \label{E:laplacian}
\end{align}
giving, for modes localised in the bulk and thus experiencing $h \sim 1$,
\begin{align}
	m_{KK} = \langle e^{\Omega(x)}\rangle \frac{1}{R_{CY}} = \frac{2\pi}{\langle\V\rangle^{1/6}}m_s \approx \frac{2\pi g_s}{\sqrt{4\pi}\V^{2/3}} M_p \,.
	\label{eq:Mkk_bulk}
\end{align}
Here, we have used $\langle e^{\Omega(x)}\rangle = 1$ and identified $R_{CY}\sim c^{1/4} l_s$ as the characteristic scale of the bulk, by analogy with a toroidal compact space, such that $(2\pi R_{CY})^6=\langle\V\rangle l_s^6$. 

In the throat region, where $e^{-4A_0} \gtrsim c$, we assume that the internal metric takes the form of the warped deformed conifold.  The deformed conifold arises from considering the surfaces in $\C^4$ given by $\sum_{i=1}^4 w_i^2 = \epsilon^2$, where $\epsilon$ is a non-zero complex parameter ($\epsilon=0$ for the singular conifold), which results in the metric \cite{candelas1990conifolds,Minasian:1999tt,Aganagic:1999fe,KS2000supergravity} 
\begin{align}
    ds_{con}^2 = \frac{\epsilon^{4/3}}{2}\mathcal{K}(\eta) 
    \Bigg( \frac{1}{3\mathcal{K}^3(\eta)} \left(d\eta^2 + (g^5)^2 \right) 
    &+ \sinh^2(\eta/2) \left( (g^1)^2 + (g^2)^2 \right) \nonumber \\
    &+ \cosh^2(\eta/2) \left( (g^3)^2 + (g^4)^2 \right)
    \Bigg)\,,
    \label{eq:deformed_conifold_metric}
\end{align}
where $\mathcal{K}(\eta) = \frac{(\sinh(2\eta) - 2\eta)^{1/3}}{2^{1/3}\sinh(\eta)}$ and $g^i$ are a basis of one-forms found in \cite{candelas1990conifolds}. We see that the metric written in this basis only depends on $\eta$, the radial direction, reflecting the symmetries of the conifold. For large $\eta$, the metric (\ref{eq:deformed_conifold_metric}) approaches the singular conifold metric \cite{candelas1990conifolds,Minasian:1999tt,KS2000supergravity,KT} 
\begin{align}
    ds_{con}^2 
    &= dr_{\infty}^2 + r_{\infty}^2\left(\frac{1}{9}(g^5)^2 +\frac{1}{6}\sum_{i=1}^{4}(g^i)^2 \right),
    \quad \text{ as } \eta\to\infty
    \label{eq:metric_conifold_largetau}
\end{align}
where the coordinate $r_\infty$ is defined as $r_{\infty}^2 = \frac{3}{2^{5/3}}\epsilon^{4/3}e^{2\eta/3}$, while near the tip, where the topology is $S^2 \times S^3$ with the $S^3$ of finite size, the full 10d metric becomes \cite{KS2000supergravity}
\begin{align}
    ds_{10}^2 &= e^{2A_0}c(x)^{-1/2} g_{\mu\nu}dx^\mu dx^{\nu} + e^{-2A_0} \left(dr_0^2 + \frac{r_0^2}{8}d\Omega_{S^2}^2 + R_\epsilon^2 d\Omega_{S^3}^2 \right), 
\end{align}
where $r_0^2 = \frac{\epsilon^{4/3}}{4}\left(\frac{2}{3}\right)^{1/3}{\eta}^2$, with the internal space of the warped metric being the deformed conifold and the warp factor given by 
\begin{align}
    e^{-4A_0(\eta)} &=
    2^{2/3}\frac{(\alpha' g_sM)^2}{\epsilon^{8/3}} I(\eta),
    \quad\quad I(\eta)\equiv \int_{\eta}^{\infty} dx ~ \frac{x\coth(x) - 1}{\sinh^2{x}}(\sinh(2x)-2x)^{1/3}\,,
	\label{eq:warp_factor}
\end{align}
where $M$ is the 3-form flux through the $S^3$ at the tip of the throat. The size of the $S^3$ in the conifold metric is controlled by the deformation parameter, $\epsilon$,
\begin{align}
    R_\epsilon^2 = \frac{\epsilon^{4/3}}{2}  \left(\frac{2}{3}\right)^{1/3}\,,
\end{align}
but its physical size at the tip of the throat, $R_{S^3}$, is independent of it due to the warp factor,
\begin{align}
    R_{S^3}^2 &= e^{-2A_0} R_\epsilon^2
    = \left(\frac{I^{1/2}(\eta)}{6^{1/3}} \right)(g_sM)\alpha' \approx (g_s M)\alpha' \,.
    \label{physsize}
\end{align}
It follows that the $\alpha'$-expansion is well under control when $g_sM\gg 1$.

Combining our discussion on the volume modulus and warped deformed conifold, we use only a finite portion of the non-compact Klebanov-Strassler solution, with the latter assumed to be glued smoothly onto a compact Calabi-Yau 3-fold.  We thus introduce a cutoff scale, $\Lambda_{UV} \sim r_{UV}$ where the conifold connects to the bulk. To this purpose, it is useful to note that, far away from the tip, the warp factor becomes \cite{KT}
\begin{align}
	e^{-4A_0} \approx \frac{L^4}{r_\infty^4}\left[1 + \frac{3 g_sM}{8\pi K} + \frac{3g_s M}{2\pi K}\log\Big(\frac{r_{\infty}}{r_{UV}}\Big) \right] \,,
	\label{E:mouthwarp}
\end{align}
where we define 
\begin{align}
	L^4 = \frac{27\pi}{4}\frac{g_sMK}{(2\pi)^4}~l_s^4 .
\end{align}
with $K$ being the flux number which appears through (\ref{eq:relation_K_Lambda0}) below. Cutting off the warped throat where the warp factor \eqref{WF} becomes $h \sim 1$ implies $r_{UV}$ is defined by $r_{UV}\sim \frac{L}{c^{1/4}}$. We define the radial size of the throat as $R_{throat}\sim L$. 

Upon dimensional reduction, the deformation parameter $\epsilon$ becomes a complex structure modulus, which is part of a chiral superfield in the 4d effective action.  It is usually denoted $|S|=\epsilon^2$ with units of $(length)^3$, or the dimensionless $z=S/l_s^{3}$.  The volume modulus $c(x) = \V^{2/3}$ also falls into a chiral superfield with scalar component $\rho$ and $c = \imrho$.  It will be useful to bear in mind the warp factor at the tip of the throat,
\begin{align}
	e^{-4A_0^{tip}} = \frac{2^{2/3}I(0)}{(2\pi)^4}\frac{(g_sM)^2}{|z|^{4/3}} \,,
	\label{eq:warp_tip}
\end{align}
and the hierarchy between the bulk (UV), where $e^{-4A_{UV}}\ll c(x)$, and the tip of the throat (IR), 
\begin{align}
	\frac{h_{IR}}{h_{UV}} = \frac{1 + \frac{e^{-4A_{0}^{tip}}}{c(x)}}{1 + \frac{e^{-4A_{UV}}}{c(x)}} \approx 1 + \frac{e^{-4A_{0}^{tip}}}{c(x)} \,,
	\label{eq:hierarchy}
\end{align}
which is large for $e^{-4A_{0}^{tip}}\gg c(x) = \V^{2/3}$.

At the tip, where the warp factor is relevant, every mass scale is suppressed. Looking at the 4d part of the metric $G_{\mu\nu} = h^{-1/2}e^{2\Omega(x)}g_{\mu\nu}$ and recalling that $\langle e^{2\Omega(x)}\rangle = 1$ by definition, energy scales measured at the tip will be suppressed compared to the bulk scales as $m^w\sim h^{-1/4} m$ \cite{quevedo2008warped}. For the warped string scale, we obtain
\begin{align}
	m_s^w = \left(\frac{e^{-4A_0^{tip}}}{c}\right)^{-1/4}m_s \,.
\end{align}
For KK modes localized near the tip, from (\ref{E:laplacian}), we derive
\begin{align}
	m_{KK}^w &= \langle e^{\Omega(x)}\rangle \frac{h^{-1/4}}{c^{1/4}}\frac{1}{R_{\epsilon}}= \frac{2\pi}{\sqrt{g_sM}} m_s = \frac{2\pi}{\sqrt{g_sM}} \frac{e^{-A_0^{tip}}}{c^{1/4}}m_s^w \,.
\end{align}

\subsection{K\"ahler potential and superpotential}
\label{S:KW}

The complete K\"ahler potential for a type IIB flux compactification can be written as \cite{giddings2003scales}
\begin{align}
    \mathcal{K}/M_p^2 =& 
    \underbrace{-3\log\left(\V^{2/3} \times\frac{\int d^6y\sqrt{g_6}~h}{V_{6}}\right)}_\text{Universal volume modulus}
    \underbrace{-\log\left(-i(\tau-\Bar{\tau})\right)}_\text{Dilaton} \nonumber \\
    &\underbrace{-\log\left(\frac{i}{\kappa_4^6}\int h~\Omega\wedge\Bar{\Omega}\right)}_\text{Complex structure moduli}
    \underbrace{-\log\left(\frac{1}{\kappa_4^6}\int d^6y\sqrt{g_6}~h\right)}_\text{Other Kahler moduli}.
    \label{eq:KahlerPotential1}
\end{align}
 For simplicity, we start with a single K\"ahler modulus, the universal volume modulus, $\rho$, with $\imrho = c = \V^{2/3}$.  Here and below we keep $V_{6} = l_s^6$ explicit for clarity, and notice the correction in the kinetic term of the volume, which is shown in \cite{frey2009universal} to arise in a warped compactification.\footnote{In equation (5.10)  of \cite{frey2009universal}, the correction to the K\"ahler potential arising from the warping is given as $\mathcal{K}=-3\log\left(V^{2/3} +\frac{\mathbf{V}_W^0}{V_{6}}\right)$, where $\mathbf{V}_W^0\equiv \int d^6y\sqrt{g_6}e^{-4A_0}$. To match (\ref{eq:KahlerPotential1}) and (5.10) of \cite{frey2009universal}, notice that 
\begin{align}
	\frac{\int d^6y\sqrt{g_6}h}{V_{6}} = \frac{\int d^6 y\sqrt{g_6}\left(1+\frac{e^{-4A_0}}{c(x)}\right)}{V_{6}} = \frac{\int d^6 y\sqrt{g_6} + \int d^6 y\sqrt{g_6}\frac{e^{-4A_0}}{c(x)}}{V_{6}}  = 1 + \frac{\mathbf{V}_W^0}{c(x)V_{6}}.
\end{align}
}
Under the usual assumption that the bulk dominates the volume of the compact space, we can neglect this correction in the volume modulus kinetic term and instead use the usual form,
\begin{align}
    \mathcal{K}/M_p^2 \approx -2\log\V.
\end{align}

\ni The metric in the complex structure moduli space can be computed using \cite{giddings2006dynamics}
\begin{align}
    G_{\alpha\Bar{\beta}} = \frac{\int h ~ \chi_\alpha\wedge\chi_{\Bar{\beta}}}{\int h ~ \Omega\wedge\Bar{\Omega}}\,.
    \label{eq:CS_moduli_metric}
\end{align}
Following closely \cite{douglas2007warping}, we review this computation for the conifold deformation modulus, $z$, using the KS metric in Appendix \ref{sec:complexstructuremoduli}, making explicit the appearance of the volume modulus. The final result, valid when $|z|\ll\Lambda_{0}^3$, is
\begin{align}
        G_{z\Bar{z}}
    = \frac{l_s^6}{\pi||\Omega||^2V_{6}} 
    \left(\log\frac{\Lambda_{0}^3}{|z|}
    + \frac{c' }{(2\pi)^4}\frac{1}{\V^{2/3}} \frac{(g_sM)^2}{|z|^{4/3}}
    \right)\,,
    \label{eq:appendixC_GSS}
\end{align}
where we define the dimensionless $\Lambda_0 = \Lambda_{UV}/l_s$ and the constant $c'=1.18$, and recall that $||\Omega||^2 = \frac{1}{3!}\Omega_{mnp}\Omega^{mnp} = 8$ is fixed by the normalisation of the globally defined covariant spinor which is a requirement for preserving $\mathcal{N}=1$ supersymmetry in 4d \cite{Lust:2004ig}.  This metric corresponds to a contribution to the K\"ahler potential of the form:
\begin{align}
    \mathcal{K}(z,\Bar{z}) =\frac{l_s^6
    }{\pi||\Omega||^2V_{6}} \left[|z|^2\left(\log\frac{\Lambda_0^3}{|z|} + 1\right) + \frac{9c'(g_sM)^2}{(2\pi)^4\V^{2/3}}|z|^{2/3}\right]\,. \label{E:K}
\end{align}
  Notice that the warping contribution to the K\"ahler potential mixes the deformation modulus $z$ and the volume modulus $\V$, in such a way that large volumes suppress the effect of the warping contribution. We can combine the K\"ahler potentials for the two moduli such that (see e.g. \cite{Dudas:2019pls})
\begin{align}
    \mathcal{K} = -3\log\left(\V^{2/3} - \frac{1}{(2\pi)^4}\frac{3c'(g_sM)^2}{\pi||\Omega||^2}\frac{l_s^6}{V_{6}}|z|^{2/3}\right) + \dots 
\end{align}
It is interesting to note that despite the mixing between $z$ and $\V$, the no-scale structure of the K\"ahler potential is preserved \cite{giddings2006dynamics,Blumenhagen:2019qcg}.

Having described the relevant terms in the K\"ahler potential, we now turn to the superpotential, which takes the well-known Gukov-Vafa-Witten form\footnote{
The normalization of the superpotential $W$ changes depending on the way we write the volume modulus and axio-dilaton terms in the K\"ahler potential. If we write $-3\log(-i(\rho-\Bar{\rho}))$, where $\imrho = c(x) = \V^{2/3}$, there is an extra factor of $2^{3/2}$ in $W$. If the Kahler potential for the axio-dilaton is written as $-\log(\Im\tau)$, there is an extra factor of $2^{-1/2}$. Similarly, if the transformation between the string and Einstein frame metrics, $G_{\mu\nu}^S = e^{\frac{\phi-\langle\phi\rangle}{2}}G_{\mu\nu}^E$, does not include the vev $\langle\phi\rangle$, the factor of $g_s$ will not be present. The same is true for $\V_w^0$, which comes from the Weyl rescaling in 4d.
} 
\begin{align}
    W =  \frac{g_s^{1/2}\sqrt{\V_w^0}}{\kappa_4^8} \int G_3 \wedge \Omega\,,
	\label{eq:GVW}
\end{align}
with the 3-form flux $G_3 = F_3 - \tau H_3$. The periods over the 3-cycles $A$ and $A^I$, $I=0,...,h^{2,1}-1$, define the coordinates in the complex structure moduli space $\int_A \Omega = S$ and $\int_{A^I} \Omega = Z^I$, and the periods over the remaining elements of the basis of 3-cycles, $B$ and $B^I$, must be functions of these coordinates. In particular \cite{douglas2007warping, upliftingrunaways2019}
\begin{align}
    \int_B \Omega = \frac{\partial F}{\partial S} = \Pi_0 + \frac{S}{2\pi i}\left(\log\frac{\Lambda_{UV}^3}{S} +1 \right) \,,  && \int_{B^I} \Omega = \frac{\partial F}{\partial Z^I}\,,
\end{align}
where $F$ is the prepotential and the period over $B$ follows from monodromy arguments for a conifold singularity while  $\Pi_0$ is independent of $S$. Given $\alpha_i,\beta^i$, the dual cohomology basis to the 3-cycles such that $\int_{A^j}\alpha_i = \int \alpha_i \wedge \beta^j = \delta_i^j$ and $\int_{B_j} \beta^i = \int \beta^i\wedge\alpha_j = -\delta^i_j$, we can write the 3-form fluxes on the 3-cycles as 
\begin{align}
    \frac{1}{(2\pi)^2\alpha'}F_3 &= M\alpha + M^0\alpha_0 - M_i\beta^i \,,\\
    \frac{1}{(2\pi)^2\alpha'}H_3 &= -K\beta - K_0\beta^0 + K^i\alpha_i \,,
\end{align}
where the RR-flux on the $S^3$ cycle at the tip of the throat, $M$, and its NSNS partner, $K$, were singled out, being the fluxes responsible for the deformation of the conifold
\begin{align}
    \frac{1}{(2\pi)^2\alpha'}\int_A F_3 
    &= \frac{1}{(2\pi)^2\alpha'}\int_{S^3} F_3
    = M \,, &
    \frac{1}{(2\pi)^2\alpha'}\int_B H_3 &=
    \frac{1}{(2\pi)^2\alpha'}\int_{\eta\leq\eta_{\Lambda}}\int_{S^2} H_3
    = K \,,
\end{align}
with $\eta_{\Lambda}$ corresponding to the radial coordinate where we cut the throat and glue it to the compact Calabi-Yau. The second integral can be computed for the conifold, using the approximation of the metric (\ref{eq:deformed_conifold_metric}) in the limit $\eta\to\infty$, (\ref{eq:metric_conifold_largetau}), with $r_{\infty}^2 = \frac{3}{2^{5/3}}\epsilon^{4/3}e^{2\eta/3}$ and the functions (\ref{eq:ansatz_functions_KS}) determined in \cite{KS2000supergravity}. We can introduce the radial cutoff in these coordinates as
\begin{align}
    \Lambda_0^2 = \frac{3}{2^{5/3}}|z|^{2/3}e^{2\eta_{\Lambda}/3}\,,
    \label{eq:cutoffscales_tau_Lambda}
\end{align}
where we replaced the deformation parameter $\epsilon$ by the complex structure $S$, since we are now looking at the 4d EFT, and rewrote the relation in terms of the dimensionless $\Lambda_0$ and $z$. It follows that
\begin{align}
    K &= \frac{1}{(2\pi)^2\alpha'}\int_{\eta\leq\eta_{\Lambda}}\int_{S^2} H_3
    \approx \frac{g_s M}{2\pi}\eta_{\Lambda}
    = \frac{g_s M}{2\pi}\Big(\log\frac{\Lambda_0^3}{|z|} + \frac{3}{2}\log\frac{2^{5/3}}{3}\Big)\,,
    \label{eq:relation_K_Lambda0}
\end{align}
which therefore must be imposed as a consistency condition for the solution $|z|$ together with the parameters $g_s,M,K,\Lambda_0$. This shows that the flux number $K$ is not an independent parameter as it might seem, but instead is related to the parameters $g_s,M$ and $\Lambda_0$, which is nothing but a reflection of the fact that the warped conifold solution \cite{KS2000supergravity} only has a free flux number $M$, since it is a solution for constant dilaton and therefore satisfies  the relation $g_s^2|F_3|^2=|H_3|^2$. Interestingly, this relation between $K$ and the cutoff scale $\Lambda_0$ takes the form of the supersymmetric strong warping solution of the deformed conifold \cite{GKP,douglas2007warping} 
\begin{align}
    |z| \approx \Lambda_0^3 e^{-\frac{2\pi K}{g_sM}}\,.
\end{align}

Finally, the Gukov-Vafa-Witten superpotential (\ref{eq:GVW}) takes the form 
\begin{align}
    W/M_p^3 
	&= \frac{g_s^{1/2}\sqrt{\V_w^0}}{\kappa_4^5}(2\pi)^2\alpha'\Big[-M\Pi_0 - \frac{M}{2\pi i}S\left(\log\frac{\Lambda_{UV}^3}{S} + 1\right) - \tau KS - M^0F_0 \nonumber \\
	&\hspace{7cm} + M_iZ^i - \tau(K_0Z^0 - K^iF_i)\Big] \\
    	&=  g_s^{1/2}\sqrt{\V_w^0}\Big(\frac{l_s}{\kappa_4}\Big)^5\left[W_0e^{i\sigma}-\frac{M}{2\pi i}z\left(\log\frac{\Lambda_0^3}{z} + 1\right) - i\frac{K}{g_s}z\right] \,.
\label{E:W}
\end{align}
We will focus on the deformation modulus $z$ and assume that the other complex structure moduli and the dilaton are stabilised by the remaining fluxes, with $\langle\tau\rangle = i g_s^{-1}$, contributing only with a constant superpotential $W_0 e^{i\sigma}$, with phase $\sigma$, corresponding to the remaining terms evaluated at their vevs. 

\subsection{The $\antiD$-brane uplift}

\label{sec:brane_potential}

In uplifting scenarios like KKLT and LVS, an $\antiD$-brane is placed at the tip of a warped throat in order to uplift the potential from an AdS minimum for the volume modulus into de Sitter\footnote{See \cite{Linde:2020mdk} for an extended version of the KKLT scenario without a supersymmetric AdS vacuum.}. This will affect the potential for the deformation modulus when the latter is lighter than the other complex structure moduli, since it is the deformation modulus that determines the warping which suppresses the brane potential \cite{upliftingrunaways2019}. The potential is obtained from the brane action in the warped background
\begin{align}
    S_{\antiD} = S_{DBI} + S_{CS} =  -2 T_3 \int d^4x \sqrt{-\det G_{\mu\nu}} \,,
   \end{align}
where $G_{\mu\nu} = h^{-1/2}e^{2\Omega(x)}g_{\mu\nu}$ is the Einstein frame metric, giving
\begin{align}
    S_{\antiD} 
    &= 2T_3 \int d^4x \sqrt{-\det g_ {\mu\nu}} \left(\frac{V_w^0}{V_w}\right)^2h^{-1}\,.   
\end{align}
Since we will be placing the $\antiD$-brane at the tip of the throat, $e^{-4A_0(\eta_{brane})}\gtrsim c(x)$ and therefore $h\approx \frac{e^{-4A_0(\eta_{brane})}}{c(x)}$. The potential is minimized for $\eta_{brane}=0$. When integrated over the whole compact space, we assume that the warping is negligible, so that $V_w\approx c(x)^{3/2}V_{6}$. Using $T_3 = \frac{1}{(2\pi)^3g_s\alpha'^2}$, the relation between $m_s$ and $M_p$ (\ref{eq:msMp}), and the warp factor at the tip (\ref{eq:warp_tip}), we obtain
\begin{align}
    V_{\antiD} = \left(\frac{g_s^3}{8\pi}\right)\frac{(2\pi)^4}{\V^{4/3}}c''\frac{|z|^{4/3}}{(g_sM)^2} M_p^4\,,
    \label{eq:D3brane_potential_Mp}
\end{align}
where we have introduced $c'' = \frac{2^{1/3}}{I(0)} \approx 1.75$.  Note that the $\antiD$-brane can equivalently be described in a supersymmetric way within the low energy effective supergravity theory using constrained superfields \cite{Kallosh:2014wsa,Bergshoeff:2015jxa,Kallosh:2015nia,Garcia-Etxebarria:2015lif,Dasgupta:2016prs,Vercnocke:2016fbt,Kallosh:2016aep, Aalsma:2017ulu, GarciadelMoral:2017vnz, Cribiori:2019hod}, though this is not necessary in what follows.

%% file: Sections/DeformationModulusStabilisation.tex
\section{Deformation Modulus Stabilisation}
\label{DeformStab_sec}
In this section we  compute the scalar potential that stabilises the deformation modulus, $z$, including the contribution from an $\overline{\text{D3}}$-brane at the tip of the KS throat.  We  identify a new regime of parameter space that occurs at large volumes and weak warping, and thus find a new minimum for the deformation modulus.  In the next section we  extend this analysis to include volume modulus stabilisation using the Large Volume Scenario \cite{LV1,originalLVS}, and find a new metastable de Sitter minimum.

\subsection{The scalar potential}
Combining  the K\"ahler potential and superpotential presented above, (\ref{E:K}) and (\ref{E:W}), the resulting scalar potential is:
\begin{align}
    V_{KS} &= \frac{g_s^3}{8}\frac{g_s}{\V^2}
   \left(\log\frac{\Lambda_0^3}{|z|} + \frac{1}{(2\pi)^4}\frac{c'(g_sM)^2}{\V^{2/3}|z|^{4/3}}\right)^{-1}\left|\frac{M}{2\pi}\log\frac{\Lambda_0^3}{z}-\frac{K}{g_s}\right|^2 M_p^4\,.
   \label{eq:secPotential_conifoldPotential}
\end{align}
In addition to this scalar potential originating from the fluxes, we include the contribution from a probe $\overline{\text{D3}}$-brane at the tip of the warped throat (in such a way that its energy contribution to the potential is suppressed down from its natural string scale)  
\begin{align}
    V_{\antiD} &= \left(\frac{g_s^3}{8\pi}\right) \frac{(2\pi)^4}{\V^{4/3}} c''\frac{|z|^{4/3}}{(g_s M)^2} M_p^4\,.
    \label{eq:secPotential_branePotential}
\end{align}
The deformation modulus appears in the brane potential through the warp factor of the metric, and we see that the suppression is provided by the vev of this modulus, through $|z|^{4/3}$. In the classic KKLT scenario \cite{KKLT}, this energy suppression  ensures that the positive energy density from the probe D-brane  uplifts an otherwise AdS minimum for the volume modulus to a near Minkowki minimum, instead of dominating the potential and causing a runaway.  How much suppression is required, and hence how large the hierarchy (\ref{eq:hierarchy}) needs to be, depends on the stabilisation mechanism of the volume modulus and, in particular, on the depth of the AdS minimum prior to the uplift. 

It is useful to introduce the following constants
\begin{align}
    \varepsilon = \frac{g_s M}{2\pi K}\,, && 
    \delta_1 = \frac{g_s^3}{8}\times\frac{K^2}{g_s} \,,&&
    \delta_2 = \frac{g_s^3}{8\pi} \times c'' \frac{c'}{\delta_1} = \frac{1}{\pi}\times c'' c' \frac{g_s}{K^2} \,,
\end{align}
as well as the parameter 
\begin{align}
    \beta \equiv \frac{\V^{2/3}\log\frac{\Lambda_0^3}{\zeta}}{\frac{c'}{(2\pi)^4}\frac{(g_sM)^2}{\zeta^{4/3}}} 
	= C \, \V^{2/3}\Lambda_0^4 ~xe^{-\frac{4}{3}x}, \label{E:beta}
\end{align}
where we  defined $z=\zeta e^{i\theta}$, introduced  the constant $C=\frac{(2\pi)^4}{c'(g_sM)^2}$, and the  useful variable $x\equiv\log\frac{\Lambda_0^3}{\zeta}$.
Using these parameters and constants,  the potentials  (\ref{eq:secPotential_conifoldPotential}) and (\ref{eq:secPotential_branePotential}) become (in Planck units,  $M_p=1$)
\begin{align}
    V &= V_{KS} + V_{\antiD} 
    = \frac{\delta_1\,C\,\,\Lambda_0^4}{\V^{4/3}} \,e^{-\frac{4}{3} x}\left[(1+\beta)^{-1}\Big(1-\varepsilon x \Big)^2 + \delta_2\right].
    \label{eq:VKSpVD3}
\end{align}
where we assumed that the axion is stabilised at zero, $\langle\theta\rangle=0$ (this will be confirmed below). 

The parameter $\beta$ measures the suppression of the warping contribution to the potential, where large values of $\beta$ reflect a large suppression of the warping. This is what we want to explore, since all previous works have assumed the regime $\beta\ll 1$, where the warping completely dominates (this is even true in the LVS case studied in \cite{originalLVS,LVSdS:2010.15903}, where the large volume of the bulk is still not relevant for the stabilisation of the deformation modulus). From the definition (\ref{E:beta}),
\be
\beta \approx \frac{\V^{2/3}}{e^{-4A_0^{tip}}} \log\frac{\Lambda_0^3}{\zeta}\,,
\ee
and we see that leaving the $\beta\ll 1$ regime requires us to have weak warping, large enough volumes and long throats.

\subsection{Strongly warped scenario ($\beta \ll 1$)}
In previous studies, it has been assumed that the warping term dominates over the volume term in (\ref{E:K}), that is $\beta\ll 1$, in which case the potential becomes 
\begin{align}
    V &\approx  \frac{\delta_1\,C}{\V^{4/3}}\,\Lambda_0^4 \,e^{-\frac{4}{3}x}\left[(1-\varepsilon x)^2 + \delta_2\right] \,,
\end{align}
and hence 
\begin{align}
    V' &= \frac{\delta_1}{\V^{4/3}}C\Lambda_0^4 \frac{4}{3}e^{-\frac{4}{3}x}\left[\left(1+\delta_2 +\frac{3}{2}\varepsilon\right) - \left(2 + \frac{3}{2}\varepsilon\right)\varepsilon x + (\varepsilon x)^2 \right]\,,
\end{align}
which shows a minimum at $\zeta=0$ ($x\to\infty$) and may or may not have critical points for $\zeta>0$ ($x$ finite). Notice that, in the absence of the brane ($\delta_2=0$) one immediately obtains the GKP solution (which corresponds to $\varepsilon x = 1$). Once the brane is introduced ($\delta_2\neq 0$), the condition that guarantees the existence of a non-trivial minimum is
\begin{align}\label{boundM}
    \frac{\delta_2}{\varepsilon^2} = \frac{4\pi c'c''}{g_sM^2} \leq \frac{9}{16},
\end{align}
and once this is satisfied we obtain the solutions
\begin{align}
    \zeta_{min} = \Lambda_0^3 ~\exp\left\{-\frac{2\pi K}{g_sM} - \left(\frac{3}{4}\pm \sqrt{\frac{9}{16}-\frac{4\pi c'c''}{g_sM^2}}\right) \right\} \,. 
\label{eq:smallbeta_zetasolution}
\end{align}
The bound \eqref{boundM} corresponds to that found in \cite{upliftingrunaways2019}, $\sqrt{g_s}M \gtrsim 6.8$, where the conclusion is that, together with the requirement that the suppression to the brane potential, through the deformation modulus
\begin{align}
    |z|^{4/3} \sim \exp\left\{-\frac{8\pi K}{3 g_s M}\right\} = \exp\left\{-\frac{8\pi (MK)}{3 g_s M^2}\right\} \,,
\end{align}
be large enough\footnote{As we mentioned before, in this argument, ``large enough'' can only be made precise when one includes the stabilisation of the volume modulus explicitly.}, this translates into a lower bound on $MK$ which makes the D3-tadpole cancellation condition difficult to satisfy (see \cite{upliftingrunaways2019,tadpoleProblem,LVSdS:2010.15903,Bena:2021tadpole}). 

\begin{figure}
    \centering
    \includegraphics[width=0.75\textwidth]{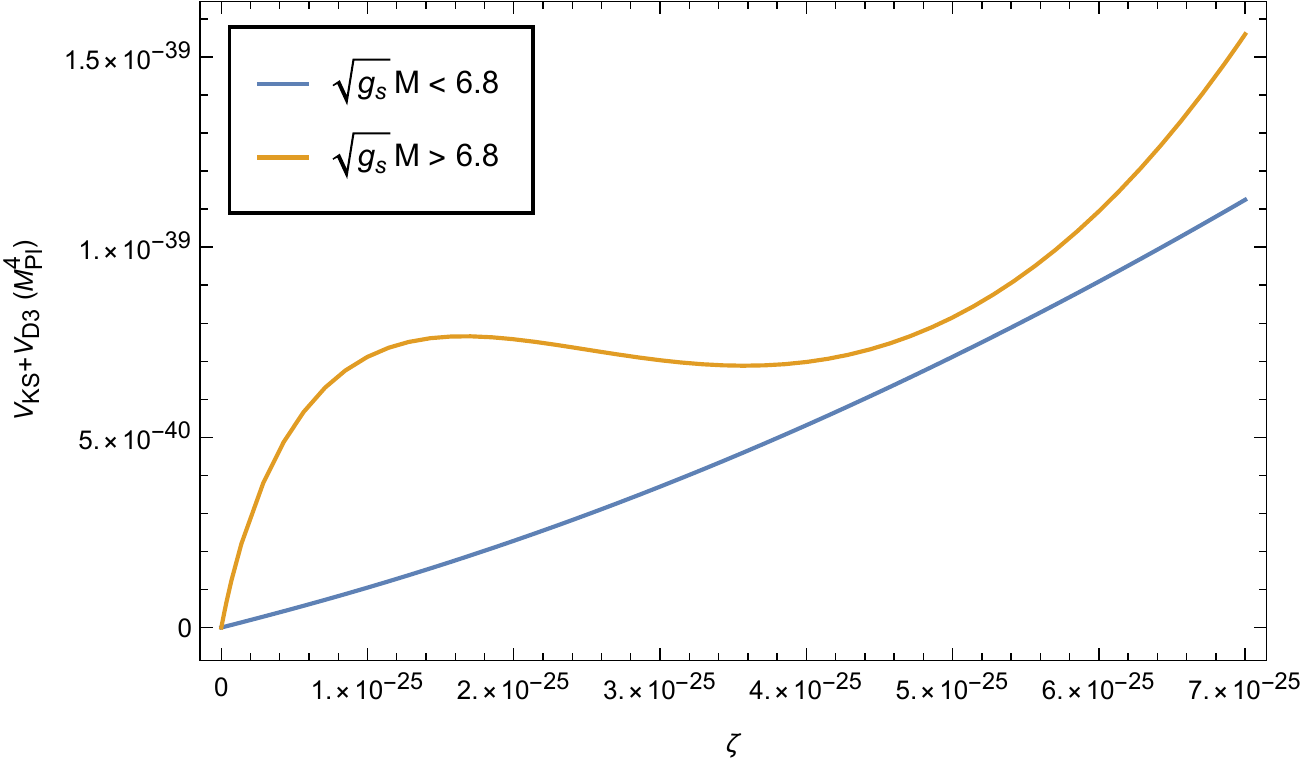}
    \caption{Comparison between two different choices of $M$ ($M=5,M=25$ with $K=M$ in both cases) for the potential (\ref{eq:VKSpVD3}) for an example with $\beta\ll 1$, with the choice of parameters $\Lambda_0=10,~g_s=0.1,~\V=10^3$ and given $||\Omega||^2 = 8,~c'=1.18,~c''=1.75$.}
    \label{fig:uplifting_original_newpotential}
\end{figure}

It is important to notice that the solution must be consistent with the approximation we started with, i.e.~the non-trivial solution $\zeta_{min}$ that follows from this analysis must be such that $\beta\ll 1$ given the vev for the volume once it is stabilised (which is verified for the parameters used in Fig. \ref{fig:uplifting_original_newpotential}).

\subsection{Weakly warped scenario ($\beta \gg 1$)}
We now consider a new regime, where the warping term in (\ref{E:K}) is subdominant, that is $\beta\gg 1$, so that the potential becomes
\begin{align}
    V &\approx \frac{\delta_1}{\V^2}\frac{1}{x}\left[\Big(1-\frac{1}{\beta}\Big)\Big(1-\varepsilon x\Big)^2 + \beta\delta_2\right] \nonumber \\
    &= \frac{\delta_1}{\V^2}\frac{1}{x}\left[\Big(1-\frac{e^{\frac{4}{3}x}}{C\,\V^{2/3}\Lambda_0^4x}\Big)\Big(1-\varepsilon x\Big)^2 + \delta_2\, C\,\V^{2/3}\Lambda_0^4\, x\,e^{-\frac{4}{3}x}\right] 
	\label{eq:}
\end{align}
and thus
\begin{align}\label{Vp}
    V' =& -\frac{2(1-\varepsilon x)(3-2x(1-\varepsilon x))}{3x^3}\frac{\delta_1}{\zeta^{7/3}\,C\,\V^{8/3}}
    +\frac{4\delta_1\delta_2 \, C\, \zeta^{1/3} }{3\V^{4/3}} 
    +\frac{3(1-(\varepsilon x)^2)}{3 x^2}\frac{\delta_1}{\zeta\V^2}\,.
\end{align}
Without further assumptions, it is difficult to solve $V'=0$ analytically, but we can still learn about the solutions by looking at different regimes for $\varepsilon x$. 

We first consider the case without an anti-D3-brane, i.e. $\delta_2=0$.  Apart from the solution at $\zeta=0$ (which one sees from (\ref{eq:secPotential_conifoldPotential}) regardless of $\beta$), we identify from (\ref{Vp})
a minimum $\zeta=\Lambda_0^3 e^{-\frac{2\pi K}{g_s M}}$ (corresponding to $\varepsilon x=1$). Given these two minima, we expect to see a maximum between them, with $\varepsilon x > 1$, which we find at $\zeta^{4/3} = \frac{c'(g_sM)^2}{(2\pi)^4}\frac{4}{3\V^{2/3}}$, differing from the maximum one finds when $\beta\ll 1$.

We now add the antibrane, i.e. $\delta_2\neq 0$. We immediately see that $\varepsilon x=1$ is not a solution of (\ref{Vp}) and, contrary to the $\beta\ll 1$ case, there is no solution near it. For $\varepsilon x \ll 1$ and large $\beta$, \eqref{Vp} becomes 
\begin{align}
    V' \approx -\frac{2(3-2x)}{3x^3}\frac{\delta_1}{\zeta^{7/3}C\,\V^{8/3}}
    +\frac{4\delta_1\,\delta_2\, C \,\zeta^{1/3} }{3\V^{4/3}} 
    +\frac{\delta_1}{x^2\zeta\V^2}\,,
\end{align}
for which again there is no solution. Finally, the case $\varepsilon x\gg 1$ implies
\begin{align}
    V' \approx \frac{\delta_1}{\V^{4/3}}\frac{4\xi}{3\zeta^{7/3}}\left(\delta_2\zeta^{8/3} - \frac{3}{4}\frac{\varepsilon^2}{C\,\V^{2/3}} \zeta^{4/3} + \frac{\varepsilon^2}{C^2\V^{4/3}}\right)\,,
\end{align}
which has solutions 
\begin{align}
    \frac{(2\pi)^4}{c'}\frac{\zeta^{4/3}}{(g_sM)^2} &=
    \frac{g_sM^2}{4\pi c'c''\V^{2/3}}\left(\frac{3}{8} \pm \sqrt{\frac{9}{64} -\frac{4\pi c'c''}{g_sM^2}}\right)\,,
    \label{eq:solution_volumedominatedregime}
\end{align}
provided that the new bound  
\begin{align}
    \sqrt{g_s}M> \frac{16}{3}\sqrt{\pi c' c''} \approx 13.6,
	\label{eq:newbound}
\end{align}
is satisfied. Notice that the maximum stays near the maximum without the brane, but the minimum no longer corresponds to a small perturbation of the GKP solution.

\begin{figure}[h!]
     \centering
     \begin{subfigure}[b]{0.48\textwidth}
         \centering
         \includegraphics[width=\textwidth]{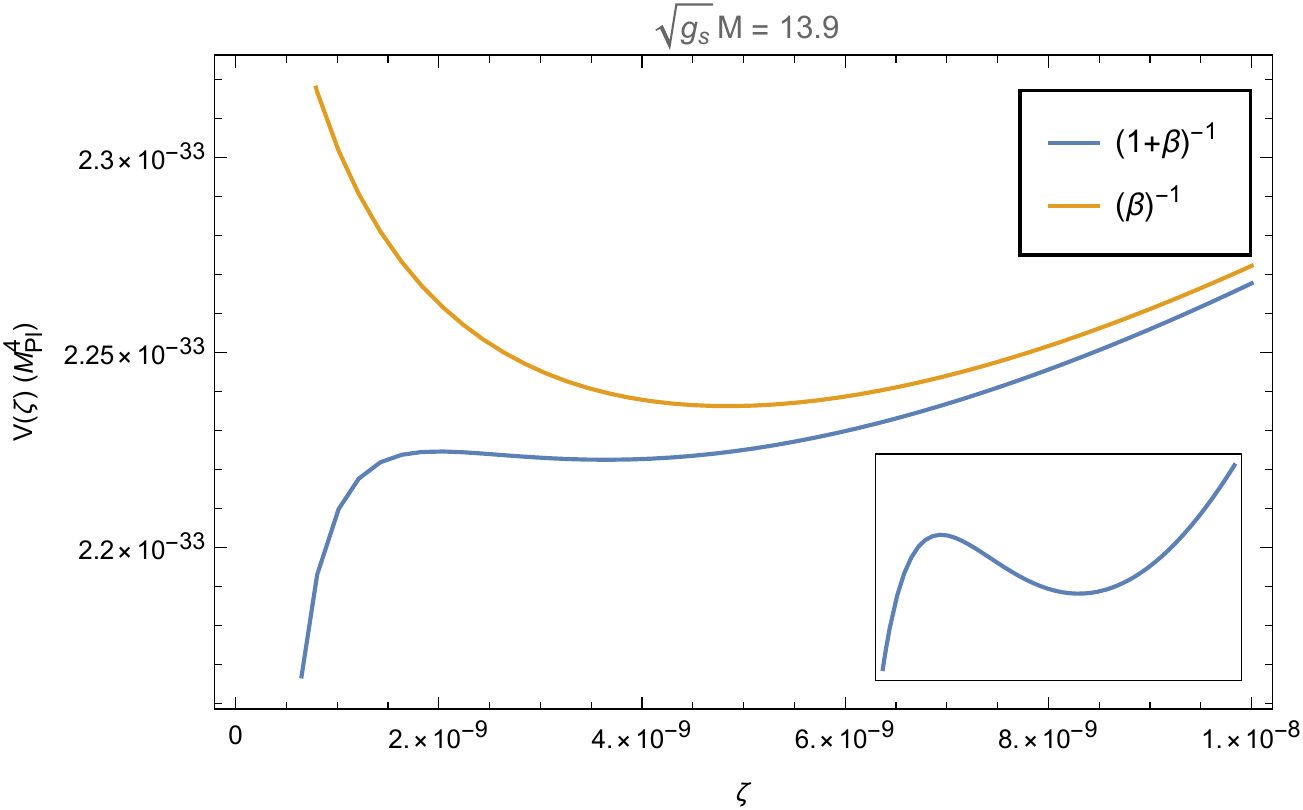}
         \label{fig:caseA}
     \end{subfigure}
     \hfill
     \begin{subfigure}[b]{0.48\textwidth}
         \centering
         \includegraphics[width=\textwidth]{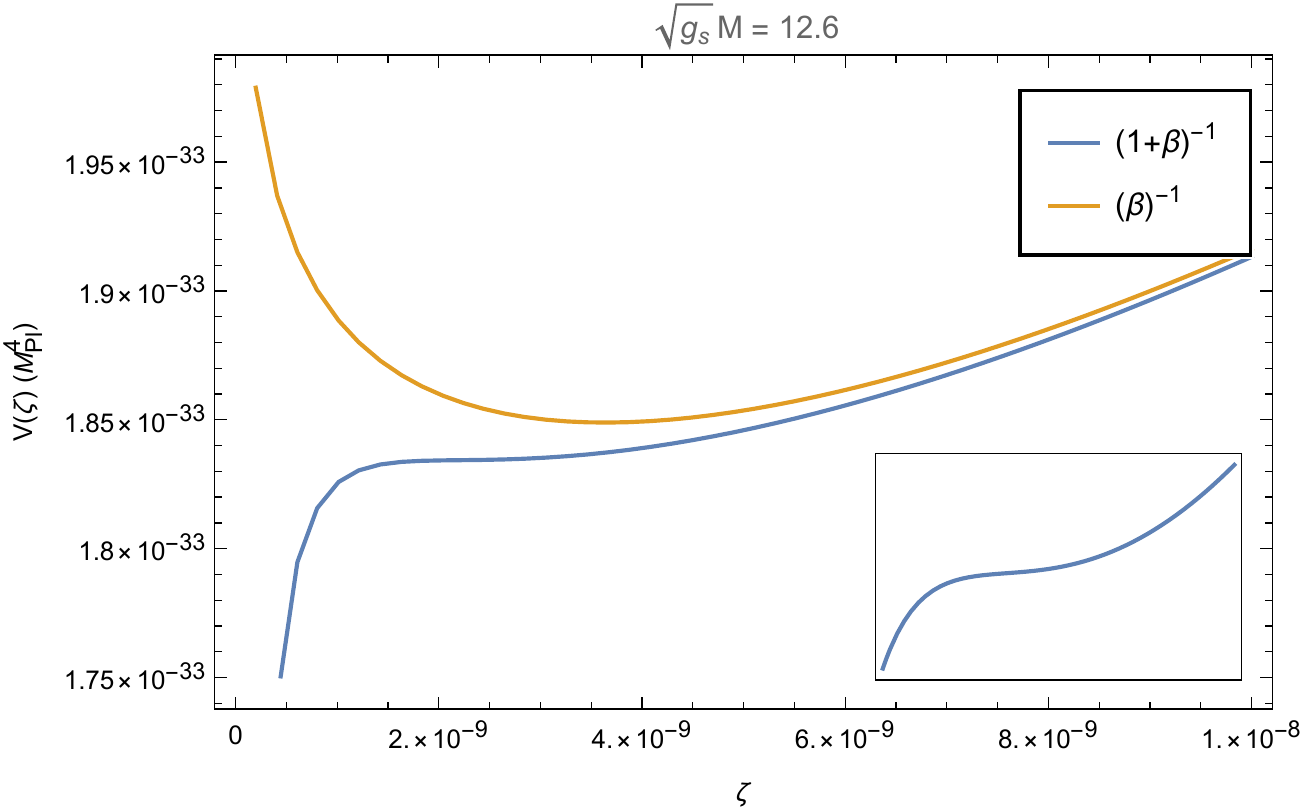}
         \label{fig:caseB}
     \end{subfigure}
        \caption{Plotting the potential (\ref{eq:VKSpVD3}) for two different choices of the flux $M$, with all other parameters fixed at $\Omega^2=8,\Lambda_0=10^2,K=1,g_s=0.1,\V=10^{15}$, in a regime with $\beta\gg 1$.  Note that we have to keep two terms in the expansion of $(1+\beta)^{-1}$ since keeping only the first term won't give us the correct behaviour as $\zeta\to 0$.  In the insets, we have zoomed in to highlight how the minimum exists or disappears depending on whether or not the bound (\ref{eq:newbound}) is satisfied.}
        \label{fig:largevolume_beta_expansion}
\end{figure}

In (\ref{eq:solution_volumedominatedregime}), $\zeta^{4/3}\sim\V^{-2/3}$, which means that the brane potential suppression comes from the volume modulus itself, provided there is a solution. This might be especially useful for the tadpole problem discussed in \cite{tadpoleProblem,Bena:2021tadpole}, since we no longer require a large hierarchy or large $MK$ (in contrast to the case $\beta\ll 1$ (\ref{eq:smallbeta_zetasolution})).  Indeed, notice that the hierarchy as defined in (\ref{eq:hierarchy}) becomes
\begin{align}
	\frac{h_{IR}}{h_{UV}}  \approx 1 + \frac{8\pi}{q_1 (g_sM^2)} \lesssim 1.14\,, \quad\text{with}\quad q_1 = \frac{3}{8} + \sqrt{\frac{9}{64} -\frac{4\pi c'c''}{g_sM^2}}\,,
\end{align}
which reflects the fact that we have a weakly-but-still-warped throat.

Given our approximate solution (\ref{eq:solution_volumedominatedregime}), we need to verify our starting assumptions. In particular, we need to check that $\beta\gg 1$ and $\varepsilon x \gg 1$.  These conditions evaluated at the solution (\ref{eq:solution_volumedominatedregime}) imply
\begin{align}
    \beta 
    &\approx \frac{3MK}{16c'c''}(\varepsilon x) \gg 1\,, \\
    \varepsilon x 
    &\approx \frac{3}{8\pi}\frac{g_sM^2}{(MK)}
    \log\left(\frac{32\pi c''\V^{2/3}}{3g_sM^2}\Lambda_0^4\right) \gg 1.
    \label{eq:conditions_newregime}
\end{align}
These can be achieved as long as 
\begin{align}
    \frac{32\pi c''}{3(g_sM^2)}\V^{2/3}\Lambda_0^4 \gg \exp{\frac{8\pi}{3}\frac{MK}{g_sM^2}}, 
\end{align}
for which we need exponentially large volumes and/or throats, with smaller values for $K$ being favourable, since we must satisfy the bound $\sqrt{g_s}M>13.6$ in order to have a solution (\ref{eq:solution_volumedominatedregime}). Given the necessity of having a large volume, we are naturally led to the Large Volume Scenario \cite{LV1,originalLVS} in an attempt to treat the volume modulus stabilisation explicitly. For comparison, we will use a similar construction to the one in \cite{LVSdS:2010.15903} and take the same ``Swiss cheese'' Calabi-Yau manifold for the bulk geometry as the one in the example therein. 

%% file: Sections/LVS.tex
\section{Volume and Deformation Moduli Stabilisation }
\label{LVS_section}
We now study the full moduli stabilisation problem including the coupled system of K\"ahler and complex structure moduli, by embedding the above system into the Large Volume Scenario.  We therefore extend our bulk K\"ahler moduli field content to include the two moduli of a ``Swiss cheese'' Calabi-Yau.  After working out the full scalar potential for this six real field system, we proceed to find metastable de Sitter vacua in the two regimes explored above, where the warping dominates or is suppressed in the deformation modulus dynamics.  The strongly-warped regime ($\beta\ll1$) was studied recently in \cite{LVSdS:2010.15903}, and we reproduce their results giving some further consistency checks.  The weakly-warped regime ($\beta\gg1$)  is new as discussed above.

\subsection{The scalar potential}

In the Large Volume Scenario \cite{originalLVS}, the K\"ahler moduli are stabilised by balancing different quantum corrections against each other. It relies on both perturbative $\alpha'$ corrections to the K\"ahler potential and non-perturbative corrections to the superpotential,
\begin{align}
    \mathcal{K}/M_p^2 &= -2\log\Big[\mathcal{V}+\frac{\xi}{g_s^{3/2}}\Big]\,, \\
    W &= W^{cs} + \sum_i A_i e^{-a_i T_i}\,,
\end{align}
where $\xi = \frac{\chi(X_6)\zeta(3)}{4(2\pi)^3}$, $\chi(X_6)$ being the Euler characteristic of the Calabi-Yau 3-fold $X_6$ that describes the internal space and $\zeta(3)\approx 1.202$, $W^{cs}$ is the flux superpotential responsible for stabilising the complex structure moduli and the dilaton, and the real part of $T_i = \tau_i + i\theta_i$ is the volume of an internal four-cycle, being therefore related to the volume of the internal space, while the axion $\theta_i$ corresponds to deformations of the RR-form $C_4$. As in \cite{LVSdS:2010.15903}, we consider a CY 3-fold with $h^{1,1}=2$ and a ``Swiss cheese'' form for the volume 
\begin{align}
    \mathcal{V} = \kappa_b \tau_b^{3/2} - \kappa_s \tau_s^{3/2}\,,
\end{align}
with $\kappa_b\equiv 1$ in what follows, which can be achieved through a suitable redefinition of $\tau_b$. We focus on the simple case where only the leading non-perturbative effect is considered \cite{LV1,originalLVS,LVSdS:2010.15903}
\begin{align}
    W = W^{cs} + Ae^{-aT_s}\,.
\end{align}

Therefore, we consider the following K\"ahler and super potentials 
\begin{align}
    \mathcal{K}/M_p^2 =& k_0 -2\log\left[\mathcal{V} + \frac{\xi}{g_s^{3/2}}\right] 
    -\log\left(-i(\tau-\Bar{\tau})\right) + k_1|z|^2\left(\log\frac{\Lambda_0^3}{|z|} + 1\right)
    +\frac{k_2}{\mathcal{V}^{2/3}} |z|^{2/3} \,, \\
    W/M_p^3 =& w_1\left(W_0e^{i\sigma} + \left[-\frac{M}{2\pi i}z\left(\log\frac{\Lambda_0^3}{z} + 1\right) - i \frac{K}{g_s} z \right] 
    + Ae^{-aT_s}\right)\,,
	\label{eq:LVS_Kahler_W}
\end{align}
with $z=\zeta e^{i\theta}$ and the parameters $k_0$, $k_1$, $k_2$ and $w_1$ defined as
\begin{align}
    k_0 &= -\log\left(\frac{||\Omega||^2V_{6}}{\kappa_4^6}\right)
	-\log\left(\frac{V_{6}}{\kappa_4^6}\right)\,, && 
    k_1 = \frac{1}{\pi||\Omega||^2}\,, &&
    k_2 = \frac{1}{(2\pi)^4}\frac{9c'(g_s M)^2}{\pi||\Omega||^2}\,, \nn \\ 
    w_1 &= g_s^{1/2}\sqrt{\V_w^0}\left(\frac{l_s}{\kappa_4}\right)^5.
\end{align}
Below we will use the expression for the gravitino mass that follows from these definitions
\begin{align}
	m_{3/2} = e^{\mathcal{K}/2}|W| \approx \frac{g_s^2W_0}{\sqrt{8\pi}\Omega\V} M_p\,. 
	\label{eq:mgravitino_bulk}
\end{align}

We now compute the scalar potential $V$ in the limit $\mathcal{V}\gg 1$ (i.e. we use the supergravity formula for the scalar potential and expand it around $1/\mathcal{V}=0$) and $\zeta\ll 1$:\footnote{We treat the term $K^{z\bar{z}}(D_zW)(D_{\bar{z}}\bar{W})$ separately, since there is a competition between these two limits, whose result depends on the $\beta$-regime we are working in. Usually, it is assumed that the warp factor completely dominates the bracket ($\beta \ll 1$) in the second line, which is equivalent to taking the limit with the constraint $\zeta^{4/3}\mathcal{V}^{2/3}\ll 1$.}
\begin{align}
    V =& \frac{g_s^3}{8\pi||\Omega||^2}
    \Bigg(\frac{8g_sa^2A^2\sqrt{\tau_s}e^{-2a\tau_s}}{3\kappa_s\V}
    + \frac{4g_s aA\tau_se^{-a\tau_s}}{\V^2}W_0\cos(a\theta_s + \sigma) 
    +\frac{3\xi}{2\sqrt{g_s}\V^3}W_0^2 \nonumber \\
    &+ \frac{g_s}{\V^{4/3}}\Big(k_1\V^{2/3}\log\frac{\Lambda_0^3}{\zeta} + \frac{k_2}{9\zeta^{4/3}}\Big)^{-1}\Bigg[\frac{M^2}{(2\pi)^2}\theta^2 + \left( \frac{M}{2\pi}\log\frac{\Lambda_0^3}{\zeta} - \frac{K}{g_s}\right)^2\Bigg] \Bigg)\,.
\end{align}

Notice that the $T_b$ axion, $\theta_b$, remains a flat direction at leading order and would be stabilised by subleading non-perturbative effects. Looking at $\partial_\theta V= \partial_{\theta_s} V= 0$, we find the solutions for the remaining axions
\begin{align}
    \langle\theta\rangle = 0 \,,&&
    \langle\theta_{s}\rangle = \frac{n\pi - \sigma}{a}, \quad n\in\Z \,,
	\label{eq:axion_vevs}
\end{align}
and choose $n=1$, such that $\cos(a\theta_s + \sigma) = -1$. By inspecting the Hessian matrix in the axion directions, $\partial_i\partial_\theta V$ and $\partial_i\partial_{\theta_s} V$, where $i$ runs through all fields, we conclude that these completely decouple from the other moduli and therefore we can 
fix the axions to their minima and then analyse the 3-field system $(\V,\tau_s,s)$. In particular, the axion masses are always positive, making these stable directions. The potential then becomes
\begin{align}
    V =& \frac{g_s^3}{8\pi||\Omega||^2}
    \Bigg(
    \frac{8g_sa^2A^2\sqrt{\tau_s}e^{-2a\tau_s}}{3\kappa_s\V} 
    - \frac{4g_s aA\tau_se^{-a\tau_s}}{\V^2}W_0
    + \frac{3\xi}{2\sqrt{g_s}\V^3}W_0^2 \nonumber\\ 
    &+ \frac{\pi g_s||\Omega||^2}{c'}\frac{(2\pi)^4}{\V^{4/3}}\frac{\zeta^{4/3}}{(g_sM)^2}\Big(1+\beta\Big)^{-1}\Big( \frac{M}{2\pi}\log\frac{\Lambda_0^3}{\zeta} - \frac{K}{g_s}\Big)^2 \Bigg)
    + \frac{g_s^3}{8\pi}\frac{(2\pi)^4}{\V^{4/3}}c''\frac{\zeta^{4/3}}{(g_sM)^2}\,,
    \label{eq:LVS_full_potential}
\end{align}
where we introduced our variable $\beta$, defined in (\ref{E:beta}), and the brane potential (see section \ref{sec:brane_potential})
\begin{align}
    V_{\antiD} = \frac{g_s^3}{8\pi}\frac{(2\pi)^4}{ \V^{4/3}}c''\frac{\zeta^{4/3}}{(g_sM)^2} \,.
\end{align}

\noindent It turns out that the solution for $\tau_b$ does not depend on the choice of $\beta$ regime, giving
\begin{align}
    \V \approx \tau_b^{3/2} = \frac{3W_0\kappa_s\sqrt{\tau_s}e^{a\tau_s}}{aA}\frac{1-a\tau_s}{1-4a\tau_s}
\end{align}
in both cases. However, we see that the solution is given in terms of $\tau_s$, which means that there is an implicit dependence on the choice for $\beta$ hiding in the solution for $\tau_s$. In turn, both $\zeta$ and $\tau_s$ will have different solutions depending on the regime of $\beta$ that we look at. We now proceed to study the two regimes of strong warping, $\beta\ll 1$, and weak warping, $\beta\gg 1$.

\subsection{Strongly warped scenario ($\beta \ll 1$)}

We now review the usual limit considered in the literature, $\beta\ll 1$. In this limit,  the potential is\footnote{Notice that our potential differs from the one in \cite{LVSdS:2010.15903}, apart from the overall factor $g_s^3/8$, in 4 ways: (i) our warp factor (\ref{eq:warp_factor}) is $e^{-4A_0}\sim(g_sM)^2$ instead of $e^{-4A_0}\sim g_sM^2$, which is a consequence of our convention for the Einstein metric with the $g_s=e^{\langle\phi\rangle}$ (see footnote \ref{F:vev_shift}); (ii) we have the factor $\pi||\Omega||^2$, which is coming from taking into account other complex structure moduli in the Kahler potential and \textit{is not an overall factor} (this contribution is hinted at in \cite{LVSdS:2010.15903} in the form of $e^{K_{cs}}$ after equation (4.3)); a factor of $(2\pi)^4$ multiplying $\zeta^{4/3}$ which comes from defining $\zeta$ in units of $l_s$; and (iv) $\Lambda_0$ is explicit as opposed to the potential used in \cite{LVSdS:2010.15903}.  
There is, however, a simple way to map the two potentials. We simply remove the overall factor in (\ref{eq:potential_smallbeta}) (which does not affect the stabilisation of the moduli in any case) and perform the following combined transformation
\begin{align}
    \zeta^{4/3} \to \Bigg(\frac{(\pi||\Omega||^2)(2\pi)^4}{g_s}\Bigg)^{-1}\zeta^{4/3} \,,
    && \Lambda_0^3 \to \Bigg(\frac{(\pi||\Omega||^2)(2\pi)^4}{g_s}\Bigg)^{-3/4}\Lambda_0^3 \,.
    \label{eq:transformation_quevedo_solution}
\end{align}}
\begin{align}
    V =& \frac{g_s^3}{8\pi||\Omega||^2}\Bigg(
    \frac{8g_sa^2A^2\sqrt{\tau_s}e^{-2a\tau_s}}{3\kappa_s\V} 
    - \frac{4g_s aA\tau_se^{-a\tau_s}}{\V^2}W_0
    + \frac{3\xi}{2\sqrt{g_s}\V^3}W_0^2 \nonumber \\ 
    &+ (\pi||\Omega||^2)\frac{ g_s}{c'}\frac{(2\pi)^4}{\V^{4/3}}\frac{\zeta^{4/3}}{(g_sM)^2}\Bigg[\frac{c'c''}{\pi g_s} + \left(\frac{M}{2\pi}\log\frac{\Lambda_0^3}{\zeta} - \frac{K}{g_s}\right)^2\Bigg]\Bigg)\,,
    \label{eq:potential_smallbeta}
\end{align}
and therefore the solution is \cite{LVSdS:2010.15903}
\begin{align}
    \tau_s^{3/2} \frac{16a\tau_s(a \tau_s-1)}{(1-4a\tau_s)^2} &= \frac{\xi}{g_s^{3/2}\kappa_s} + (2\pi)^4(\pi||\Omega||^2)\frac{8q_0 \zeta^{4/3}\tau_b^{5/2}}{27 g_s^2\kappa_sW_0^2} \,, \\
    \zeta &= \Lambda_0^3~ e^{-\frac{2\pi}{g_s}\frac{K}{M} - \left(\frac{3}{4} \pm \sqrt{\frac{9}{16} - \frac{4\pi c' c''}{g_sM^2}}\right)}\,,
    \label{eq:analytical_solution_smallbeta}
\end{align}
with the constant $q_0 = \frac{3}{8\pi^2 c'}\big(\frac{3}{4} - \sqrt{\frac{9}{16} - \frac{4\pi c' c''}{g_sM^2}}\big)$.
\vspace{1mm}

The value of the potential at the critical points is given by
\begin{align}
    V_{crit} &= \frac{g_s^3}{8\pi||\Omega||^2}\left(\frac{\pi||\Omega||^2}{g_s}(2\pi)^4\frac{5q_0\zeta^{4/3}}{9\tau_b^2} - \frac{3W_0^2g_s\kappa_s\sqrt{\tau_s}}{4a\tau_b^{9/2}}\frac{16a\tau_s(a\tau_s - 1)}{(1-4a\tau_s)^2}\right) \nonumber \\
    &\approx 
    \frac{g_s^3}{8\pi||\Omega||^2}\left(\frac{\pi||\Omega||^2}{g_s}(2\pi)^4\frac{5q_0\zeta^{4/3}}{9\tau_b^2} - \frac{3W_0^2g_s\kappa_s\sqrt{\tau_s}}{4a\tau_b^{9/2}}\right),
\end{align}
which must be positive if we want to have a dS solution. Whether this corresponds to a local minimum (rather than a maximum or a saddle) is related to the masses of the three fields and is analysed in \cite{LVSdS:2010.15903}. Two of the mass-eigenvalues are always positive, but a second bound is derived from requiring 
\begin{align}
    m_3^2 \approx \frac{3}{2\tau_b^2}\left(\frac{5}{4}\frac{27W_0^2g_s\kappa_s\sqrt{\tau_s}}{20a\tau_b^{5/2}} - \frac{\pi||\Omega||^2}{g_s}(2\pi)^4 q_0\zeta^{4/3}\right)>0\,.
\end{align}
Satisfying the two conditions $V_{crit}>0$ and $m_3^2>0$ provides a constraint on the value of $\rho\equiv \frac{\pi||\Omega||^2}{g_s}(2\pi)^4 q_0\zeta^{4/3}$ \cite{LVSdS:2010.15903}, which can be written as $\alpha\in [1,5/4]$ where
\begin{align}
    \rho = \alpha  \frac{27W_0^2g_s\kappa_s\sqrt{\tau_s}}{20a\tau_b^{5/2}}.
\end{align}

The hierarchy (\ref{eq:hierarchy}) in this regime becomes
\begin{align}
	\frac{h_{IR}}{h_{UV}}  \sim 1 + \frac{e^{\frac{8\pi K}{3g_sM}}}{\Lambda_0^4 \V^{2/3}}\,,
\end{align}
which depends not only on the flux numbers and string coupling, but also on the volume $\V$ (which is now exponentially large) and the length of the throat $\Lambda_0$, large values of which will decrease the hierarchy between the UV and the IR.

\subsubsection{Example}
\begin{figure}[ht]
\begin{subfigure}{.45\textwidth}
  \centering
  \includegraphics[width=1.\linewidth]{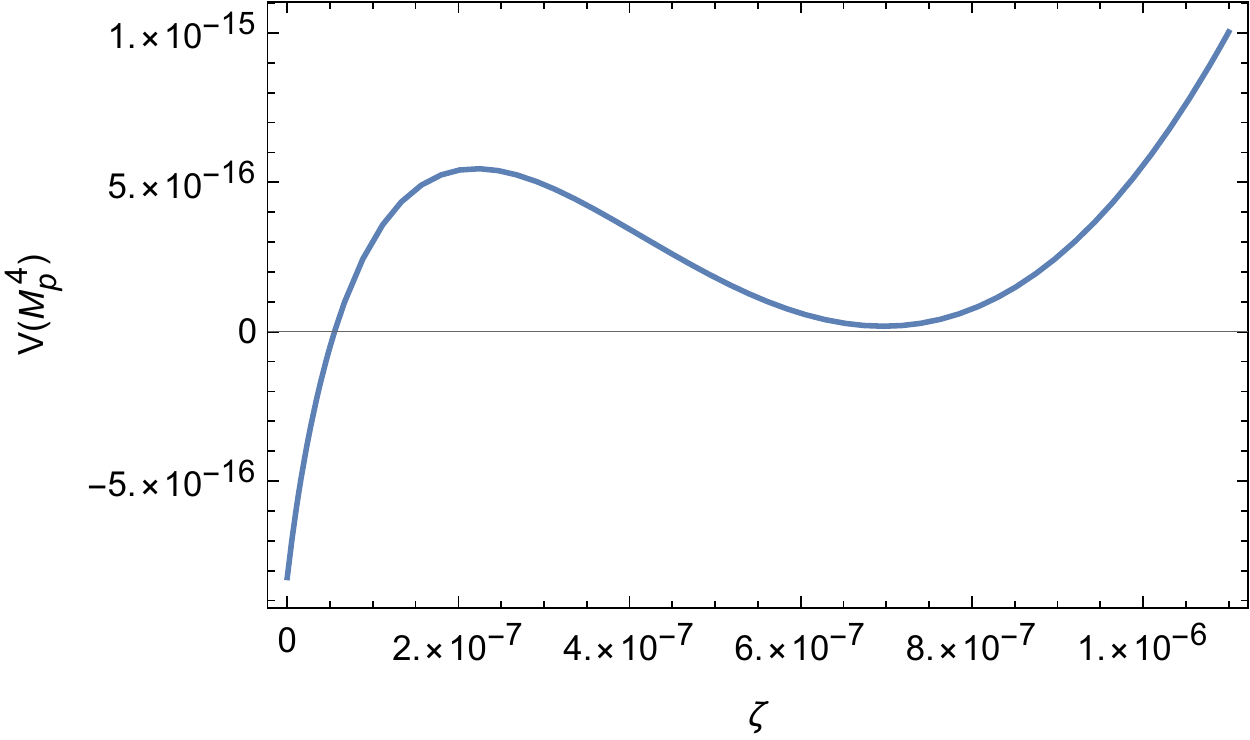}
  \label{fig:Example1_plot1_(taub,V)}
\end{subfigure}
\begin{subfigure}{.45\textwidth}
  \centering
  \includegraphics[width=1.\linewidth]{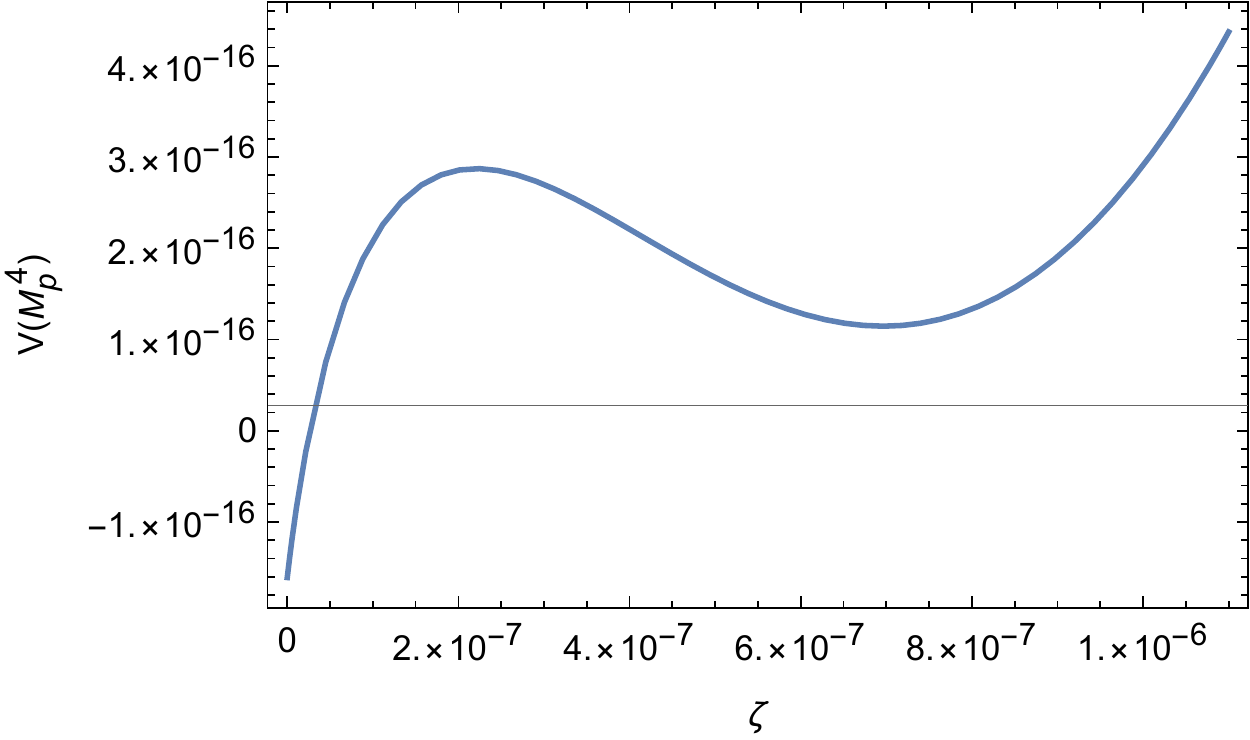}
  \label{fig:Example1_plot1_(taub,V)}
\end{subfigure}
\begin{subfigure}{.45\textwidth}
  \centering
  \includegraphics[width=1.\linewidth]{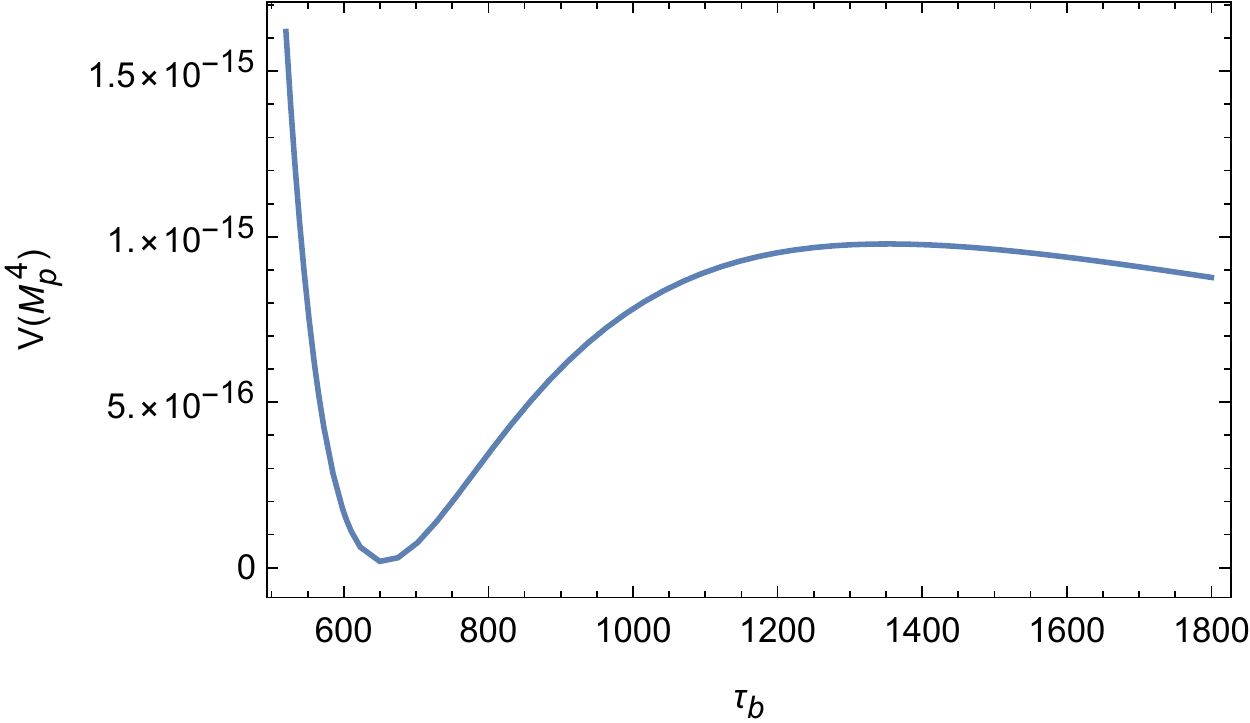}
  \label{fig:Example1_plot1_(taub,V)}
\end{subfigure}
\begin{subfigure}{.45\textwidth}
  \centering
  \includegraphics[width=1.\linewidth]{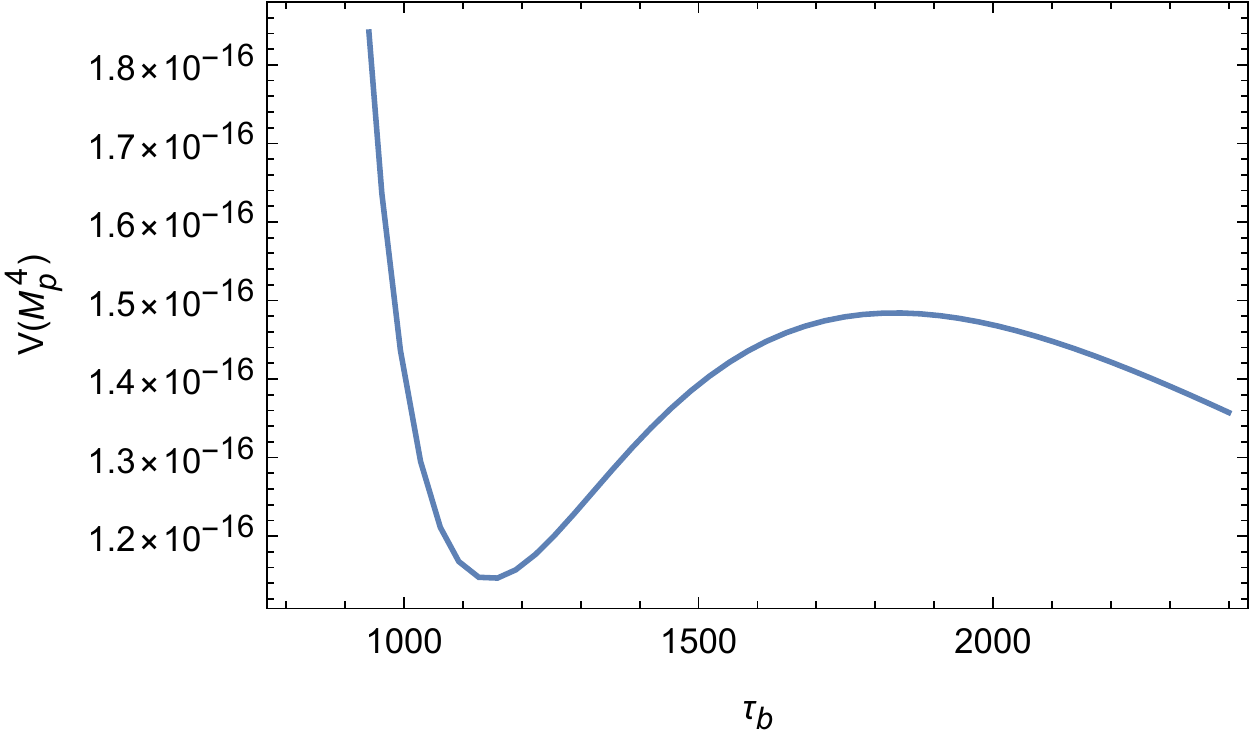}
  \label{fig:Example1_plot2_(taub,V)}
\end{subfigure}
\begin{subfigure}{.45\textwidth}
  \centering
  \includegraphics[width=1.\linewidth]{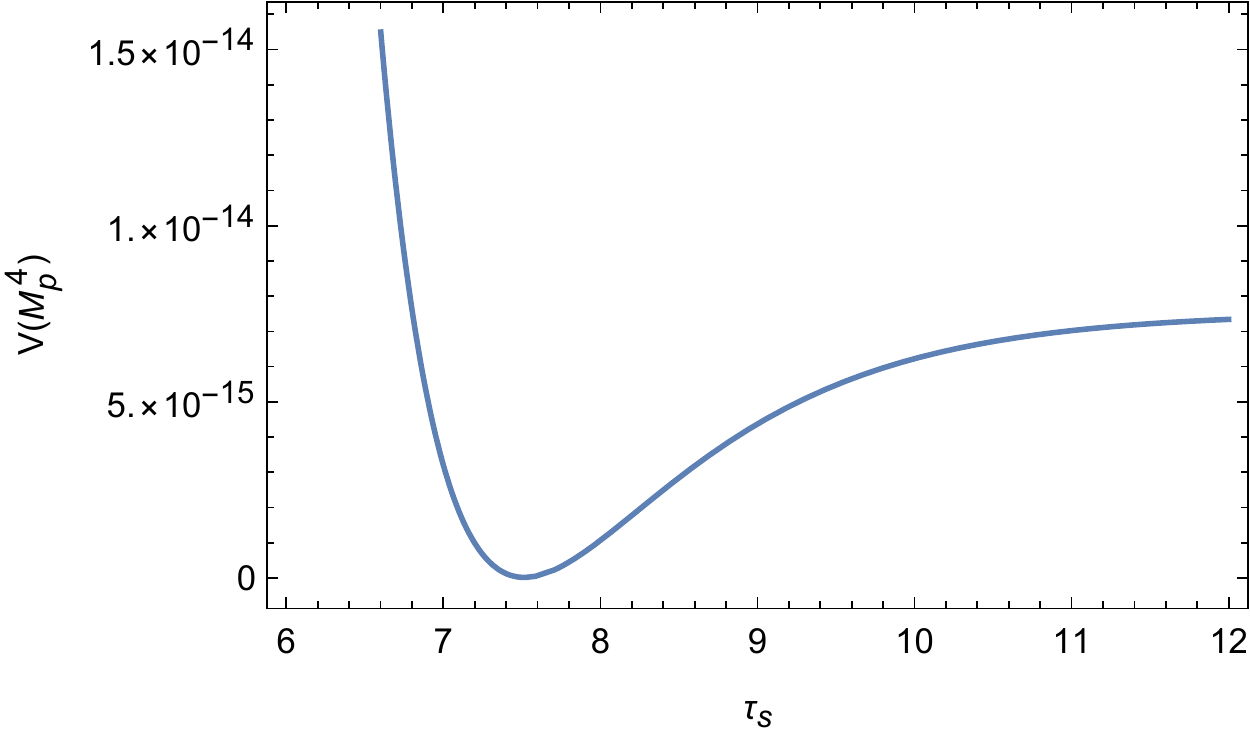}
  \label{fig:Example1_plot1_(taus,V)}
\end{subfigure}\hspace{15mm}
\begin{subfigure}{.45\textwidth}
  \centering
  \includegraphics[width=1.\linewidth]{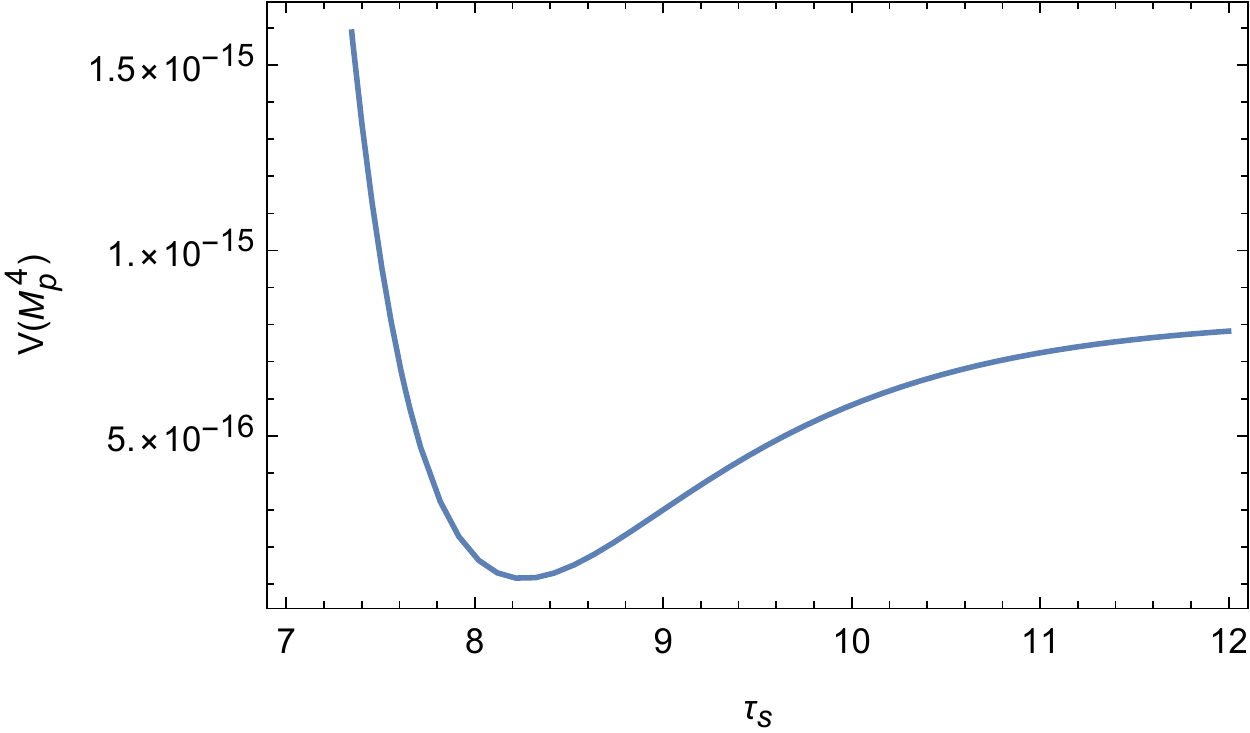}
  \label{fig:Example1_plot2_(taus,V)}
\end{subfigure}
\caption{Plots of the potential (\ref{eq:LVS_full_potential}) in each direction $(\tau_s,\tau_b,\zeta)$ near the minimum (left) and the saddle point (right), for the choice of parameters in Table \ref{Tab1}. }
\label{fig:plotsLVS_directions}
\end{figure}

\begin{figure}[ht]
	\centering
  	\includegraphics[width=.7\linewidth]{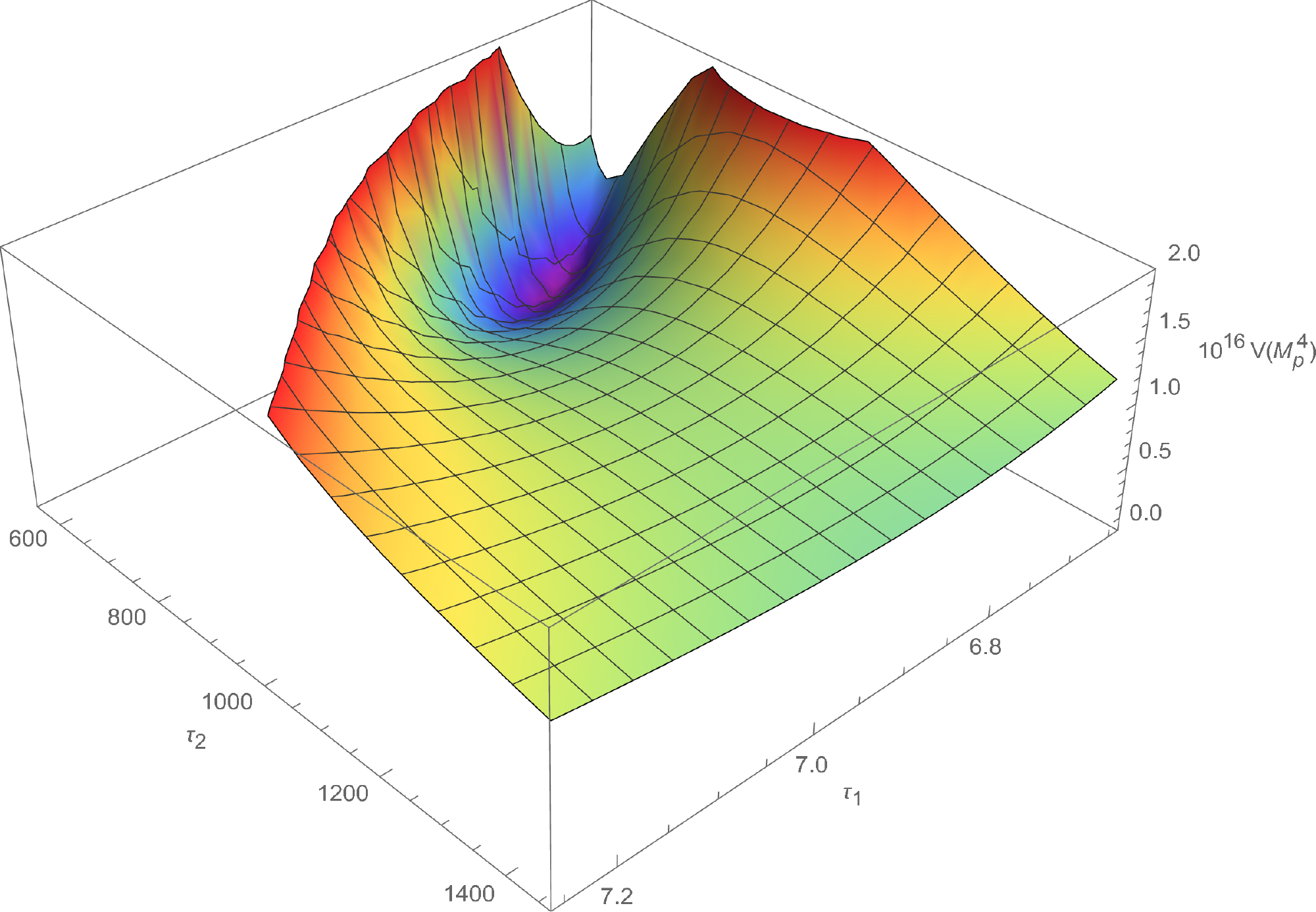}
	\caption{Plots of the potential (\ref{eq:LVS_full_potential}) in the $(\tau_1,\tau_2)$ plane (which is just a rotation of the $(\tau_b,\tau_s)$ plane aligned with the eigenvectors of the Hessian matrix), for the choice of parameters in Table \ref{Tab1}, where we can see both the minimum and saddle point solutions.}
	\label{fig:plotsLVS_directions_3d}
\end{figure}

Using a set of parameters which corresponds to the example given in \cite{LVSdS:2010.15903}, we find the expected minimum and saddle point with one unstable direction. This is summarised in Table \ref{Tab1} and Figs. \ref{fig:plotsLVS_directions} and \ref{fig:plotsLVS_directions_3d}.

\begin{table}[h!]
	 \begin{tabular}{| c | c | c | c | c | c | c | c | c | c | c | } 
	 \hline
	 \rowcolor{black!10!white!90} $W_0$ & $\sigma$ & $g_s$ & $M$ & $K$ & $\Lambda_0$ & $\kappa_s$ & $\chi$ & $a$ & $A$ \\  
	 \hline
	 $23$ & 0 & $0.23$ & $22$ & $4$ & $0.05$ & $\frac{\sqrt{2}}{9}$ & $-260$ & $\frac{\pi}{3}$ & $1$ \\ 
	 \hline
	 \end{tabular}
	\newline
	 \begin{tabular}{| c | c | c | c | c | c | c |} 
		\hline
		\rowcolor{black!10!white!90} Solution & $\tau_s$ & $\tau_b$ & $\zeta$ & $m_1^2\sim m_\zeta^2$ & $m_2^2\sim m_{\tau_s}^2$ & \multicolumn{1}{c|}{$m_3^2\sim m_{\tau_b}^2$} \\  
		\hline
		Minimum & 7.52 & 655 & $6.96\times 10^{-7}$
			& $1.31\times 10^{-5}$ & $ 2.76\times 10^{-9} $ &   \multicolumn{1}{c|}{$ 1.56\times 10^{-15}$} \\
		\hline
		Saddle & 8.26 & 1144 & $6.96\times 10^{-7}$
			& $7.51\times 10^{-6}$ & $ 6.34\times 10^{-10} $ & \multicolumn{1}{c|}{$ -2.04\times 10^{-16}$}  \\
		\hline
	 \end{tabular}
	\begin{tabular}{| c | c | c | c | c | c | c | c |}
	\hline
		\rowcolor{black!10!white!90} $\V$ & $M_s$ & $m_{KK}$ & $m_{3/2}$ & $M_s^w$ & $m_{KK}^w$  \\  
		\hline
		 $1.68\times 10^{4}$ & $3.15\times 10^{-3}$ & $6.22\times 10^{-4}$ & $5.12\times 10^{-6}$ & $3.82\times 10^{-4}$ & $1.40\times 10^{-3}$ \\ 
		\hline
		 $3.87\times 10^{4}$ & $2.07\times 10^{-3}$ & $3.56\times 10^{-4}$ & $2.22\times 10^{-6}$ & $2.89\times 10^{-4}$ & $9.21\times 10^{-4}$ \\ 
		\hline
	\end{tabular}
	\caption{Solution and masses for the fields $(\tau_s,\tau_b,\zeta)$, and physical scales associated with the solution, for a set of parameters with $\beta\ll 1$. The mass scales are expressed in units of $M_p$.}\label{Tab1}
\end{table}

This solution gives the hierarchy of scales $m_{3/2}<m_{s}^w<m_{KK}<m_{KK}^w<M_s$. The cutoff scale will be $M_s^w$,\footnote{If the masses of the other complex structure moduli and the dilaton are significantly lighter than this scale, then they should be integrated in.} with $m_{3/2} < M_s^w$ consistently with the 4d supergravity EFT description (see also \cite{Cicoli:2013swa}). Although $M_s^w < m_{KK}^w$, the $\alpha'$ expansion remains well under control, being governed by $R^2_{S^3} \ll \alpha'$ or $g_s M \gg 1$ near the tip, as can be verified by estimating the suppression of higher derivative terms in the warped background.  Two other possible control issues that have been raised in the recent literature \cite{Carta_2019,Gao:2020xqh} are the danger that the throat does not `fit' into the bulk, and that singularities are induced in the bulk with no physical interpretation.  For the present solution, $2.59\approx \pi R_{CY}>R_{throat}\approx 0.72$ so that the throat fits the bulk, where the factor of $\pi$ comes from approximating the bulk with a torus, as pointed out in Appendix A of \cite{Gao:2020xqh} and as consistent with the estimate of $m_{KK}$ made in (\ref{eq:Mkk_bulk}). Moreover, the solution, corresponding to the usual LVS, also avoids the bulk singularity problem due to the hierarchy $\tau_s\ll \tau_b\sim\V^{2/3}$, where $\tau_s$ corresponds to the small cycle wrapped by the E3 or D7 branes that are responsible for the non-perturbative contribution to the superpotential.\footnote{In detail, the bulk-singularity problem \cite{Gao:2020xqh} emerges when we consider two ways that the physics at the tip of the throat is tied to physics in the bulk region.  The first relation is from the requirement that the $\antiD$-brane energy, fixed by the warp factor and thus the flux numbers at the tip, uplifts the KKLT AdS vacuum to an almost Minkowski spacetime, that is $V_{AdS} \sim V_{uplift}$. The second relation comes from the negative D3-brane-charge sources that must be present in the bulk in order to cancel the positive D3-brane-charge sourced by the fluxes within the throat.  Together, these relations imply that the warp factor on the wrapped 4-cycle in the bulk satisfies $\frac{|\partial h|}{h} \gtrsim g_s M^2 \gg 1$ leading to a singularity where the warp factor becomes negative, apparently far from any sources that might resolve the singularity. The LVS solution circumvents this problem by relaxing the bound to $\frac{|\partial h|}{h} \gtrsim \frac{\tau_s}{\tau_b} g_s M^2$.}

Finally, we consider the consistency condition (\ref{eq:relation_K_Lambda0}), which comes from integrating the NSNS flux through the B-cycle stretching along the cutoff warped throat and corresponds to the strongly warped solution of \cite{GKP,douglas2007warping}.  The current solution is indeed in the strong warping regime and the fact that the volume is large (compared to the alternative KKLT setup) does not affect the solution for the conifold, which effectively still does not feel the effects of a large bulk. Therefore, it is not surprising that, given the analytical solution for $\zeta$ (\ref{eq:analytical_solution_smallbeta}), the relation (\ref{eq:relation_K_Lambda0}) is approximately respected, as the probe brane only brings a small correction to the supersymmetric solution with no brane \cite{GKP}.

\subsection{Weakly warped scenario ($\beta \gg 1$)}

We now consider the new limit where the warping is subdominant in the K\"ahler metric for the conifold modulus, $\beta\gg 1$. In this case, 
 the potential becomes\footnote{We keep two terms in the expansion of $(1+\beta)^{-1}$ as explained in Fig.\ref{fig:largevolume_beta_expansion}.}
\begin{align}
    V =& 
\frac{g_s^3}{8\pi||\Omega||^2}\Big(
    \frac{8g_sa^2A^2\sqrt{\tau_s}e^{-2a\tau_s}}{3\kappa_s\V} 
    - \frac{4g_s aA\tau_se^{-a\tau_s}}{\V^2}W_0
    + \frac{3\xi}{2\sqrt{g_s}\V^3}W_0^2 \Big)   \nonumber \\ 
    &+ \frac{g_s}{\V^{2}\log\frac{\Lambda_0^3}{\zeta}}\Bigg(1-\frac{\frac{1}{(2\pi)^4}\frac{c'(g_sM)^2}{\zeta^{4/3}}}{\V^{2/3}\log\frac{\Lambda_0^3}{\zeta}}\Bigg)\Big[ \frac{M}{2\pi}\log\frac{\Lambda_0^3}{\zeta} - \frac{k}{g_s}\Big]^2 
   + \frac{g_s^3}{8\pi}\frac{(2\pi)^4}{\V^{4/3}}c''\frac{\zeta^{4/3}}{(g_sM)^2}\,, \label{E:largebetaV}
\end{align}
which only has a minimum for $\zeta$ in the limit $\varepsilon\log\frac{\Lambda_0^3}{\zeta}\gg 1$, with $\varepsilon=\frac{g_s M}{2\pi K}$,
\begin{align}
    \tau_s^{3/2}\frac{16a\tau_s(a\tau_s-1)}{(1-4a\tau_s)^2} =& \frac{\xi}{g_s^{3/2}\kappa_s} \\
    + (\pi||\Omega||^2)&\tau_b^{3/2}\left(\frac{8\tau_b}{27 \pi g_s\kappa_s W_0^2}c''\frac{(2\pi)^4\zeta^{4/3}}{(g_sM)^2} + \frac{4\left(\frac{M}{2\pi}\log\left(\frac{\Lambda_0^3}{s}\right)-\frac{k}{g_s}\right)^2}{9W_0^2\kappa_s\log\left(\frac{\Lambda_0^3}{s}\right)}\right)\,, \nonumber \\
	(2\pi)^4 c''\frac{\zeta^{4/3}}{(g_sM)^2} =& \frac{g_sM^2}{4\pi \tau_b}\left(\frac{3}{8} \pm \sqrt{\frac{9}{64}-\frac{4\pi c'c''}{g_sM^2}}\right)\,,
\end{align}
from which we extract the expected bound
\begin{align}
     \sqrt{g_s}M> \frac{16}{3}\sqrt{\pi c' c''} \approx 13.6 . \label{E:newbound}
\end{align}
As we have seen with (\ref{eq:solution_volumedominatedregime}), this solution is not a perturbation of the GKP solution. In particular, the minimum is not a small uplift of a solution without the brane, but rather a new solution which relies on the presence of the brane. In particular, the brane uplifting is suppressed due to the large volume, whilst the solution always has weak warping and hence a small hierarchy of scales (\ref{eq:hierarchy})
\begin{align}
	\frac{h_{IR}}{h_{UV}} \approx 1 + \frac{8\pi}{q_1 (g_sM^2)} \lesssim 1.14\,, \quad\text{with}\quad q_1 = \frac{3}{8} + \sqrt{\frac{9}{64} -\frac{4\pi c'c''}{g_sM^2}}\,,
\end{align}
with $\frac{3}{8}<q_1<\frac{3}{4}$ following from (\ref{E:newbound}). 

\begin{figure}
    \centering
    \includegraphics[scale=0.5]{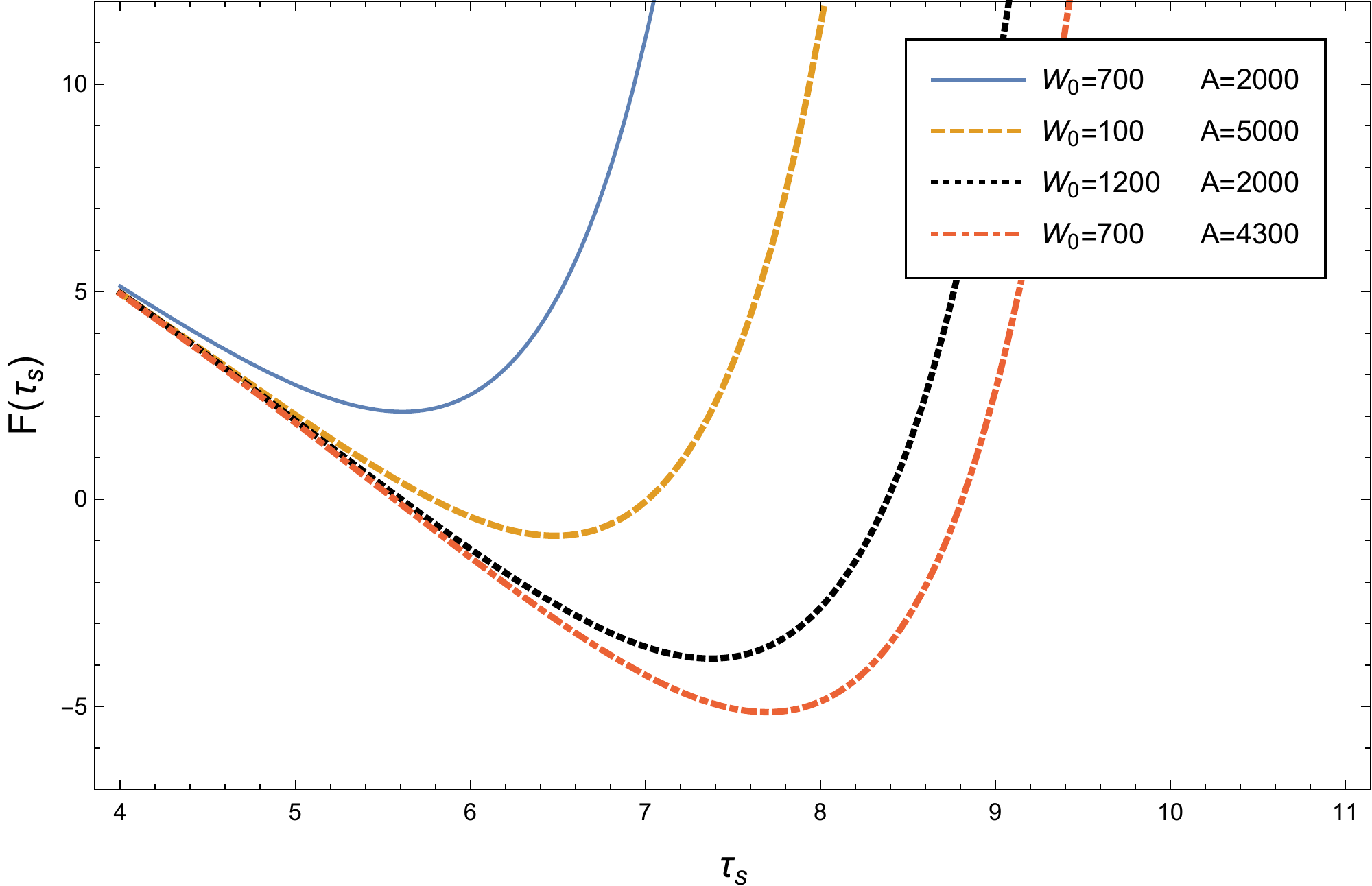}
    \caption{The plot shows (\ref{eq:Ftaus}) for different choices of the parameters $W_0$ and $A$, with all other parameters fixed. Critical points for the scalar potential (\ref{E:largebetaV}) exist when $F(\tau_s)=0$.  We see that the product $W_0A$ (which appears in the equation (\ref{eq:equation_in_terms_of_taus})) needs to be large enough for a solution to exist.}
    \label{fig:condition_for_taus_solution}
\end{figure}

It is difficult to give the solution for the fields explicitly in terms of the parameters. However, we can rewrite the equations in a form that allows us to find a numerical solution. In detail, we write
\begin{align}
    \V \approx \tau_b^{3/2} &\approx \frac{3W_0\kappa_s\sqrt{\tau_s}e^{a\tau_s}}{4aA}\,, \label{eq:solV} \\
    (2\pi)^4 c''\frac{\zeta^{4/3}}{(g_sM)^2} &= \frac{g_sM^2}{4\pi}q_1\left(\frac{3W_0\kappa_s\sqrt{\tau_s}e^{a\tau_s}}{4aA}\right)^{-2/3} \,,\\
    \tau_s^{3/2} &\approx \frac{\xi}{g_s^{3/2}\kappa_s} + (\pi||\Omega||^2)\frac{\frac{M^2}{(2\pi)^2}\log\left(\frac{\Lambda_0^3}{\zeta}\right)}{3aAW_0}e^{a\tau_s}\sqrt{\tau_s}\,,
    \label{eq:equation_in_terms_of_taus}
\end{align}
where we use $\varepsilon x\gg 1$ and $\log(\frac{\Lambda_0^3}{\zeta})\gg q_1$. In order for a solution to (\ref{eq:equation_in_terms_of_taus}) to exist, the exponential term cannot be too large since it has to balance with the $\tau_s^{3/2}$ on the opposite side of the equation. In Fig.~\ref{fig:condition_for_taus_solution}, we plot the function
\begin{align}
    F(\tau_s) = -\tau_s^{3/2} + \frac{\xi}{g_s^{3/2}\kappa_s} + (\pi||\Omega||^2)\frac{\frac{M^2}{(2\pi)^2}\log\left(\frac{\Lambda_0^3}{\zeta}\right)}{3aAW_0}e^{a\tau_s}\sqrt{\tau_s}\,,
\label{eq:Ftaus}
\end{align}
defined such that $F(\tau_s)=0$ corresponds to a solution to (\ref{eq:equation_in_terms_of_taus}), where we have replaced $\zeta$ by its solution in terms of $\tau_s$. 
In the approximation $a\tau_s\gg 1$, we can estimate a lower bound for the product $AW_0$ as 
\begin{align}
	A W_0 \gtrsim \frac{\pi||\Omega||^2}{9}\frac{M^2}{(2\pi)^2}\log\left(\frac{\Lambda_0^3}{\zeta}\right) \exp\left[\frac{a\xi^{2/3}}{g_s\kappa_s^{2/3}}\right] \,,
	\label{eq:AW0_bound}
\end{align}
by requiring the minimum of $F(\tau_s)$ to be negative. 
For a set of favourable parameters, e.g. $\xi\sim\mathcal{O}(1),~\kappa_s\sim\mathcal{O}(10^{-1}),~g_s\sim 0.2,~M\sim\mathcal{O}(10)$ as well as 
$\log\left(\frac{\Lambda_0^3}{\zeta}\right)\sim\mathcal{O}(10)$ so that $\zeta\ll 1$ with $\Lambda_0\sim\mathcal{O}(1)$, we find $AW_0\gtrsim 10^6$. This suggests that we  require large values of $W_0$ and/or $A$ in order to have a solution in this region of parameter space. Notice also that if $A$ becomes too large, $\V$ becomes small, and that large values of $AW_0$ push the minimum of the potential towards AdS (cf.~(\ref{eq:largebeta_Vmin}) below), which implies that we need to find a balance between these parameters.

The value of the potential at the critical points is now
\begin{align}
    V_{crit} 
	&\approx \frac{g_s^4}{24\tau_b^3}\left(\frac{M^2}{(2\pi)^2}\log\left(\frac{\Lambda_0^3}{\zeta}\right) - \frac{3}{\pi||\Omega||^2}(AW_0)e^{-a\tau_s}\right) \,,
    \label{eq:largebeta_Vmin}
\end{align}
which must be positive for a dS solution. In the above, we use the approximations $a\tau_s\gg 1, ~\varepsilon x \gg 1,~ q_1\ll\log\left(\frac{\Lambda_0^3}{\zeta}\right)$. In order to determine whether this critical point is a minimum, we compute the mass matrix 
\begin{align}
    \mathbf{M} = h^{ab}\frac{\partial^2V}{\partial\varphi^a\partial\varphi^b}\Big|_{crit}
\end{align}
where $\varphi^a=\{\tau_b,\tau_s,\zeta\}$ and $h_{ab}$ is the field space metric defined via $K_{i\Bar{j}}\partial\Phi^i\partial\Bar{\Phi}^{\Bar{j}} = \frac12 h_{ab} \partial \varphi^a \partial \varphi^{b}$.  
To find the masses $(m_1^2,m_2^2,m_3^2)$, we diagonalise $\mathbf{M}$ in the limit $\mathcal{V}\approx\tau_b^{3/2}\gg 1$ and using $a\tau_s\gg 1$ and $\varepsilon x\gg 1$.  We obtain, in Planck units:
\begin{align}
    m_1^2 &\approx \frac{g_s^3}{8}\frac{q_2}{9\sqrt{\pi}g_s^{3/2}\tau_b^{3/2}} \frac{(2\pi)^6 c''^{\,3/2}(\pi||\Omega||^2)}{(g_sM)^2q_1^{5/2}\log\left(\frac{\Lambda_0^3}{\zeta}\right)} \,,\\
    m_2^2 &\approx \frac{g_s^3}{8}\frac{1}{\pi||\Omega||^2}\frac{4g_s W_0^2 a^2\tau_s^2}{\tau_b^3} \,,\\
    m_3^2 &\approx \frac{3g_s^4}{8\tau_b^3}\left(\frac{3}{2}\frac{3}{\pi||\Omega||^2}(AW_0)e^{-a\tau_s} - \frac{M^2}{(2\pi)^2}\log\left(\frac{\Lambda_0^3}{\zeta}\right)\right)\,,
\end{align}
where $q_2=q_1(9+4q_1)-\frac{112\pi c'c''}{g_sM^2}$, which we can show is always positive provided there is a solution. Since $q_1>\frac{3}{8}$, indeed
\begin{align}
    q_2 > \frac{3}{8}\frac{21}{2} - \frac{112\pi c c''}{g_sM^2} = 28\left(\frac{9}{64} - \frac{4\pi c'c''}{g_sM^2}\right) > 0.
\end{align}
We also used the fact that a solution requires large enough $W_0$ to simplify the expression for $m_2^2$. Note that only $m_3^2$ can be negative and gives us a condition for the minimum. Putting this together with $V_{crit}>0$ from (\ref{eq:largebeta_Vmin}), we find
\begin{align}
   \frac{3}{\pi||\Omega||^2}(AW_0)e^{-a\tau_s} < \frac{M^2}{(2\pi)^2}\log\left(\frac{\Lambda_0^3}{\zeta}\right) < \frac{3}{2}\times\frac{3}{\pi||\Omega||^2}(AW_0)e^{-a\tau_s} \,. \label{E:newmetadScdns}
\end{align}
Defining the parameter
\begin{align}
    \mu &\equiv \frac{\frac{M^2}{(2\pi)^2}\log\left(\frac{\Lambda_0^3}{\zeta}\right)}{\frac{3}{\pi||\Omega||^2}(AW_0)e^{-a\tau_s}}  \,,
\label{eq:muparameter}
\end{align}
 the condition for a metastable dS solution is simply $1<\mu<\frac{3}{2}$. A positive potential only requires $\mu>1$, so cases with $\mu>\frac{3}{2}$ will be at best dS saddle points, which could be used for example for quintessence or inflation. Notice also that the condition for the existence of a solution, (\ref{eq:AW0_bound}), corresponds to $\mu\lesssim 3$, which is compatible with a metasable dS solution.

It is important to note that the consistency condition (\ref{eq:relation_K_Lambda0}), which comes from integrating the NSNS flux through the B-cycle that extends along the cutoff warped throat, is not automatically satisfied for our new de Sitter solution. In fact, the condition 
\begin{align}
    K  \approx \frac{g_s M}{2\pi}\log\frac{\Lambda_0^3}{\zeta}
    \implies
    1 \approx \varepsilon x 
    \label{eq:relation_K_Lambda0_largebeta}
\end{align}
is not compatible with the region where the regime $\beta \gg 1$ has a solution ($\varepsilon x \gg 1$). This reflects the fact that this regime does feel the influence of a large bulk and the solution for $\zeta$ depends explicitly on the volume. In such a case, the approximation where the B-cycle is completely contained within the warped throat is no longer good and bulk effects should become relevant and contribute to the flux. It is therefore not too surprising that the relation \eqref{eq:relation_K_Lambda0_largebeta} is farther from being satisfied in this regime, where the value $\varepsilon x$ quantifies how far way we are from satisfying it. 

Another important consistency requirement is $m_{3/2}<m_{KK}$, so that the gravitino mass of the 4d EFT remains below the cutoff and is not integrated out. Using the scales (\ref{eq:Mkk_bulk}) and (\ref{eq:mgravitino_bulk}), together with the solution (\ref{eq:solV}) and the condition (\ref{eq:AW0_bound}) required for a solution to exist, we can write the ratio
\begin{align}
	\Big(\frac{m_{3/2}}{m_{KK}}\Big)^3 = \frac{g_s^3W_0^3}{4\pi\sqrt{2}||\Omega||^3\V} \gtrsim \frac{W_0}{\sqrt{2}||\Omega|| M^{5/2}}\frac{(a\tau_s)(g_sM)^{7/2}K}{9\xi}(\varepsilon x)\,,
\end{align}
where recall that $a\tau_s\gg 1$, $g_sM\gg 1$, $\varepsilon x\gg 1$, and the fact that a  too small $W_0$ results in a small volume. This makes it difficult to find a region in parameter space for which the supergravity description remains valid, however, we show a working example below.

\subsubsection{Example}
\label{sec:examples}

We choose a set of parameters that satisfies the conditions (\ref{E:newmetadScdns}) discussed above, which guarantee both $V_{crit}>0$ and $m_3^2>0$ and thus a dS minimum. We have parameters associated with the fluxes for the remaining complex structure moduli and the axio-dilaton $(W_0,g_s)$, with the conifold $(M,K,\Lambda_0)$ and with the CY and Kahler moduli $(\kappa_s,\xi,a,A)$, where $\xi = -\frac{\chi \zeta(3)}{4(2\pi)^3}$. The fixed parameters in the potential are $c' = 1.18$, $c'' = 1.75$, $||\Omega||^2 = 8$.

In Table \ref{Tab2} we present a set of parameters that serve to provide a concrete example of the stabilisation mechanism presented in this subsection.

\begin{table}[h!]
\centering
 \begin{tabular}{| c | c | c | c | c | c | c | c | c | c | } 
 \hline
 \rowcolor{black!10!white!90}  $W_0$ & $\sigma$ & $g_s$ & $M$ & $K$ & $\Lambda_0$ & $\kappa_s$ & $\chi$ & $a$ & $A$ \\  
 \hline\hline
$700$ & $0$ & $0.14$ & $39$ & $3$ & $1.1$ & $\frac{\sqrt{2}}{9}$ & $-150$ & $\frac{\pi}{3}$ & $4300$ \\ 
 \hline
 \end{tabular}
\caption{Choice of  parameters for the potential (\ref{eq:LVS_full_potential}), with $\beta\gg 1$.
}
\label{Tab2}
\end{table}

For the parameter set in Table \ref{Tab2}, we  find two solutions for equation (\ref{eq:equation_in_terms_of_taus}) and therefore two approximate solutions. 
We solve the system numerically, using these as initial conditions, to determine the true critical points of the full potential (\ref{eq:LVS_full_potential}).  We find one minimum and one saddle point with a single unstable direction (roughly corresponding to $\tau_b$). The solutions are summarised in Table \ref{Tab3}. 
\begin{table}[h!]
\centering
	 \begin{tabular}{| c | c | c || c | c | c |} 
		\hline
		\rowcolor{black!10!white!90} $\tau_s$ & $\tau_b$ & $\zeta$ & $m_1^2\sim m_\zeta^2$ & $m_2^2\sim m_{\tau_s}^2$ & $m_3^2\sim m_{\tau_b}^2$ \\  
		\hline\hline
		8.79 & 62.8 & 0.0062 
			& $1.56\times 10^{-4}$ & $5.77\times 10^{-4}$ & $3.11\times 10^{-9}$ \\ \hline
		9.09 & 78.5 & 0.0054 
			& $1.05\times 10^{-4}$ & $3.14\times 10^{-4}$ & $-1.30\times 10^{-10}$ \\ 
		\hline
	 \end{tabular}
	\caption{Solution and masses (in units of $M_p$) for the fields $(\tau_s,\tau_b,\zeta)$ for the two parameter sets in Table \ref{Tab2}. The first solution is metastable, whereas the second has negative $m_3^2$, revealing an unstable direction.}\label{Tab3}
\end{table}

\begin{center}
\begin{table}[h!]
\centering
	 \begin{tabular}{| c | c | c | c | c | c | c | c |} 
		\hline
		\rowcolor{black!10!white!90} $\V$ & $M_s$ & $m_{KK}$ &  $m_{3/2}$ & $M_s^w$ & $m_{KK}^w$ \\  
		\hline\hline
		498 & $1.11 \times 10^{-2}$ & $3.95\times 10^{-3}$ & $1.95\times 10^{-3}$ & $1.04\times 10^{-2}$ & $4.76\times 10^{-3}$ \\ \hline
		696 & $9.41\times 10^{-3}$ & $3.16\times 10^{-3}$ & $1.39\times 10^{-3}$ & $8.82\times 10^{-3}$ & $4.03\times 10^{-3}$ \\
		\hline
	 \end{tabular}
	\caption{Physical scales associated with the solutions in Table \ref{Tab3}, for the parameter set in Table \ref{Tab2}. The mass scales are expressed in units of $M_p$}
\label{tb:physical_scales_weakwarping}
\end{table}
\end{center}

From Table \ref{tb:physical_scales_weakwarping} we see that the cutoff for the EFT is now $m_{KK}\sim m_{KK}^w$, since this is the smallest scale, which reflects the fact that this solution has weak warping.  Note that $\frac{m_{3/2}}{m_{KK}}\approx \frac12 $, which is marginally consistent with the 4d supergravity description (see also \cite{Cicoli:2013swa}). We have $1.41\approx \pi R_{CY}>R_{throat}\approx 0.69$, so that the throat still fits the bulk \cite{Carta_2019,Gao:2020xqh}.  
Similarly to the usual LVS, this solution avoids the bulk singularity problem \cite{Gao:2020xqh,Carta:2021lqg} due to the hierarchy\footnote{In detail, using (\ref{eq:largebeta_Vmin}) for the uplift to near Minkowski along with the arguments in \cite{Gao:2020xqh}, the warp factor on the wrapped 4-cycle can be shown to satisfy $\frac{|\partial h|}{h} \gtrsim \frac{\tau_s}{\tau_b}\frac{g_s}{\log{W_0 A}}$ so that bulk singularities can be avoided.} $\tau_s\ll \tau_b\sim\V^{2/3}$.
 
\begin{figure}[ht]
\begin{subfigure}{.45\textwidth}
  \centering
  \includegraphics[width=1.\linewidth]{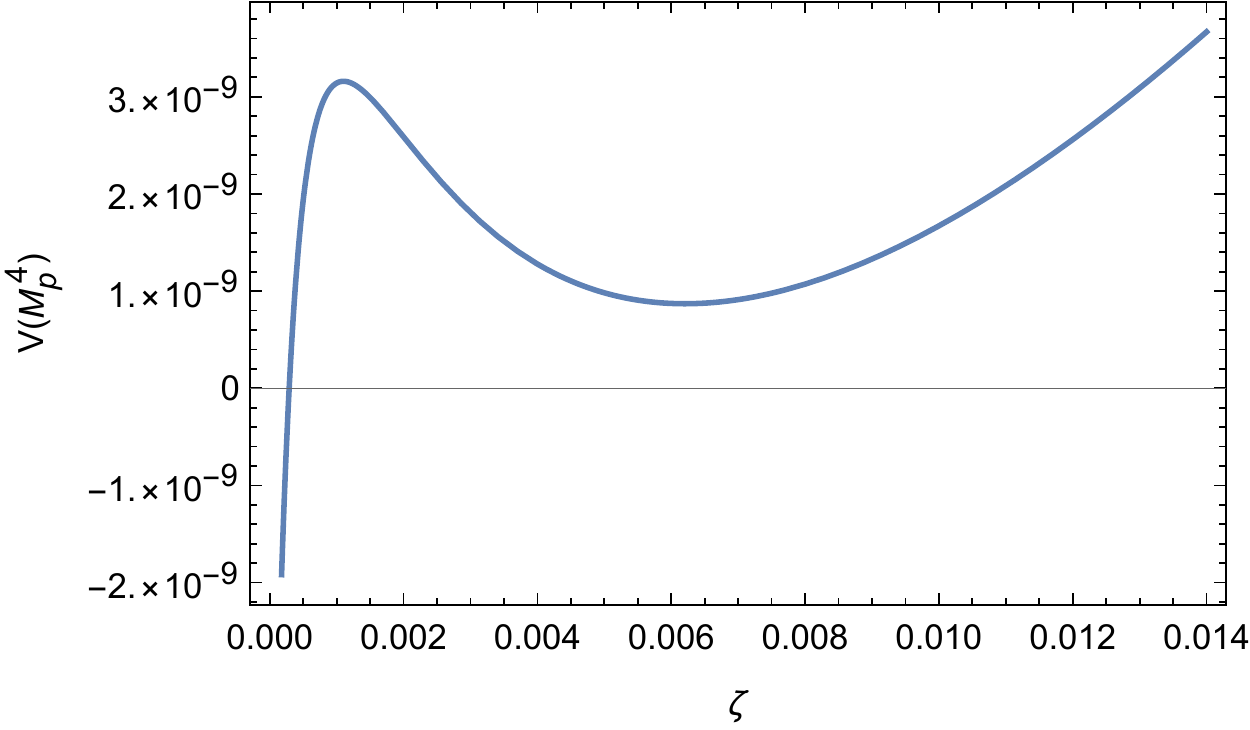}
\end{subfigure}
\begin{subfigure}{.45\textwidth}
  \centering
  \includegraphics[width=1.\linewidth]{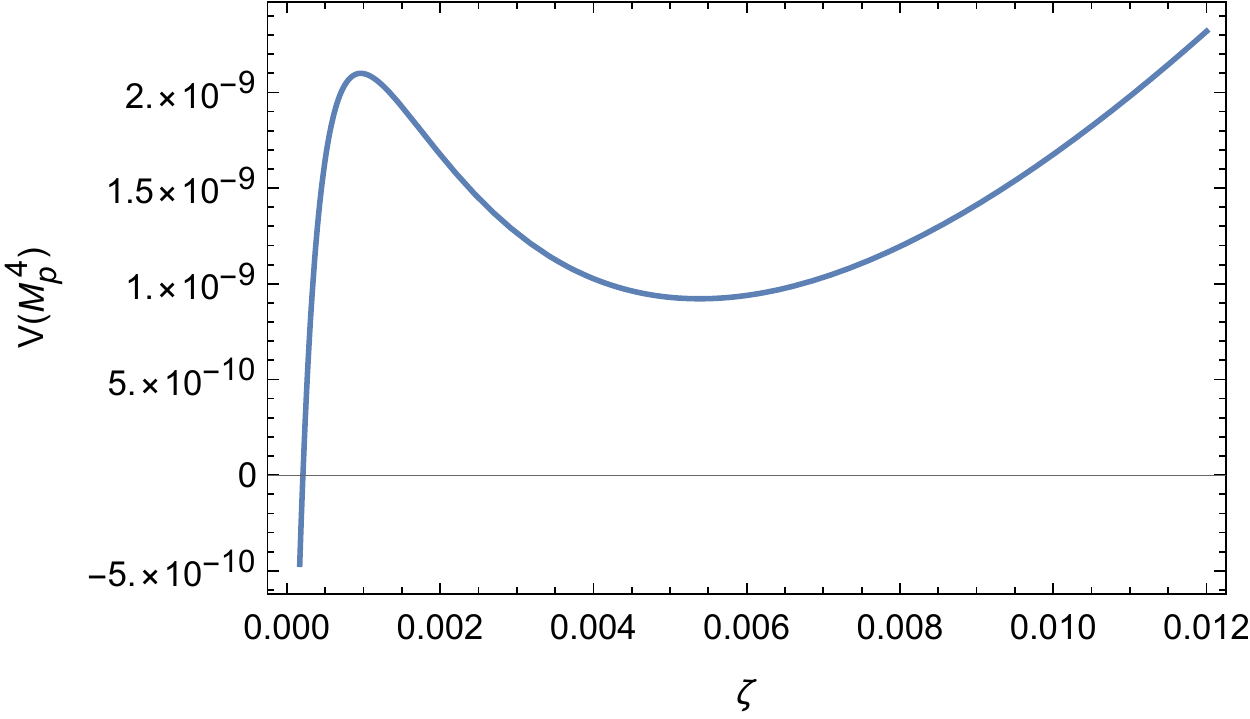}
\end{subfigure}
\begin{subfigure}{.45\textwidth}
  \centering
  \includegraphics[width=1.\linewidth]{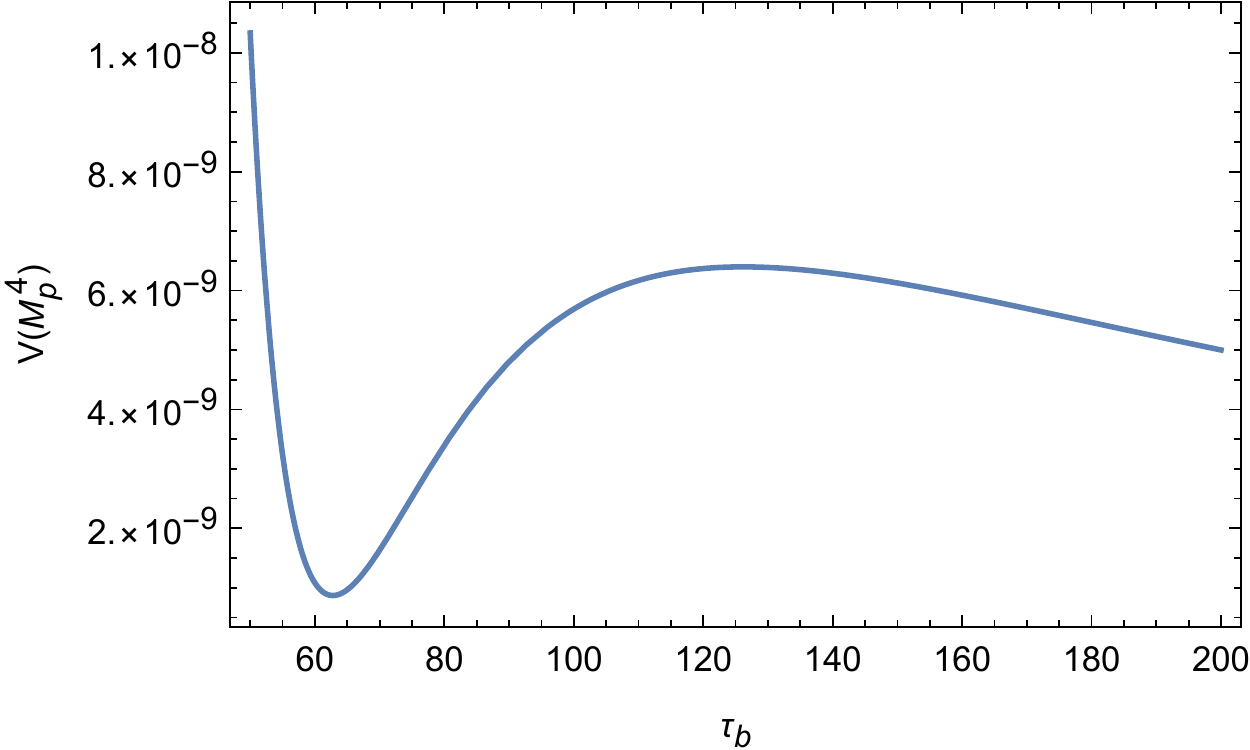}
\end{subfigure}
\begin{subfigure}{.45\textwidth}
  \centering
  \includegraphics[width=1.\linewidth]{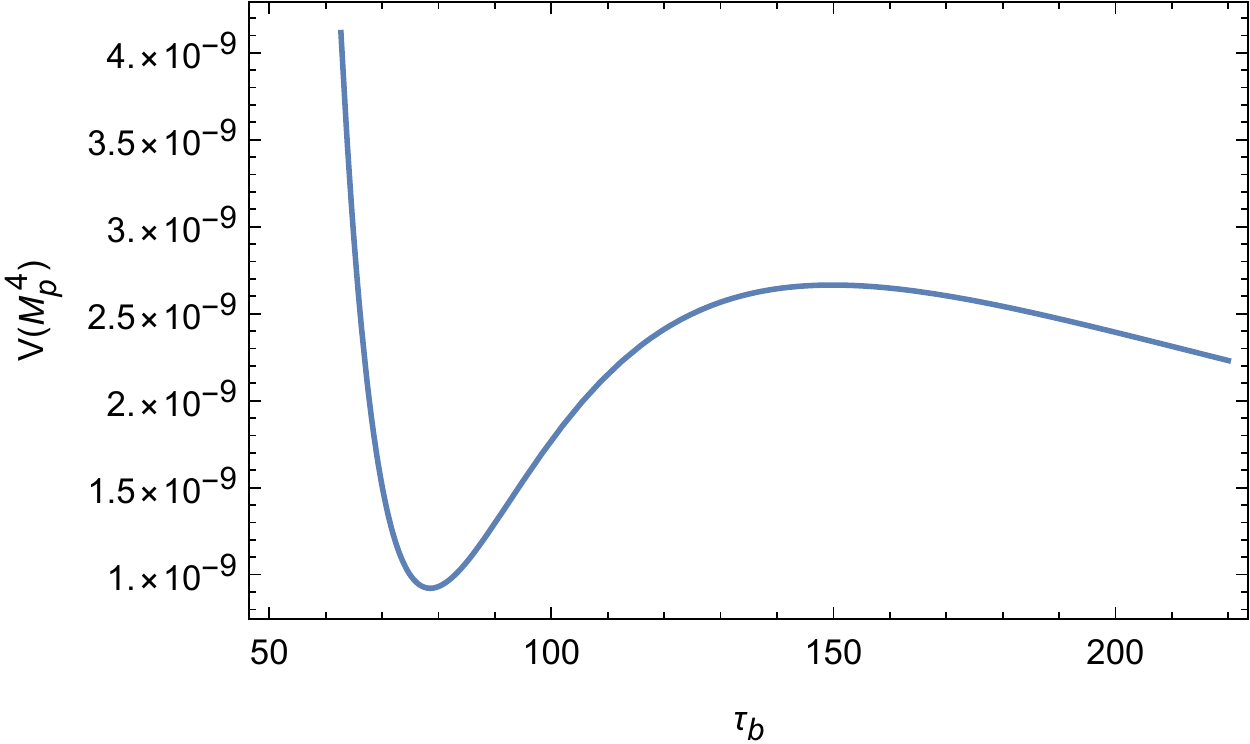}
\end{subfigure}
\begin{subfigure}{.45\textwidth}
  \centering
  \includegraphics[width=1.\linewidth]{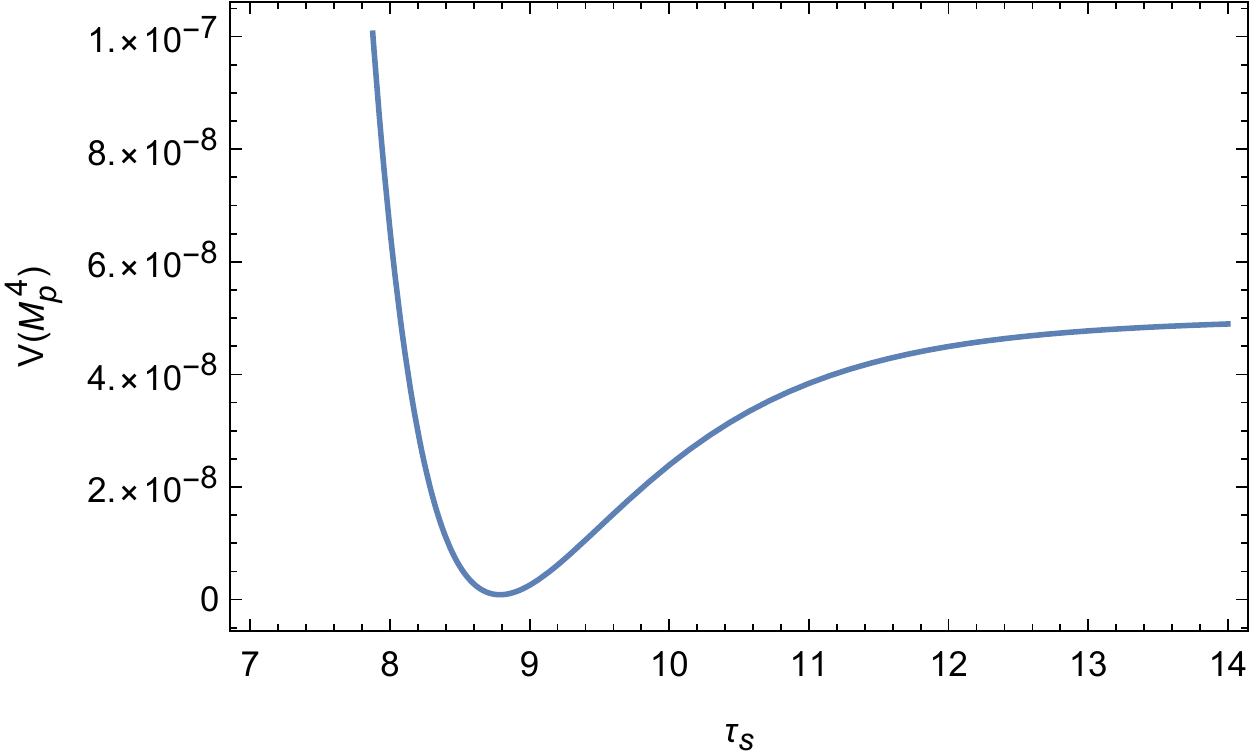}
\end{subfigure}\hspace{15mm}
\begin{subfigure}{.45\textwidth}
  \centering
  \includegraphics[width=1.\linewidth]{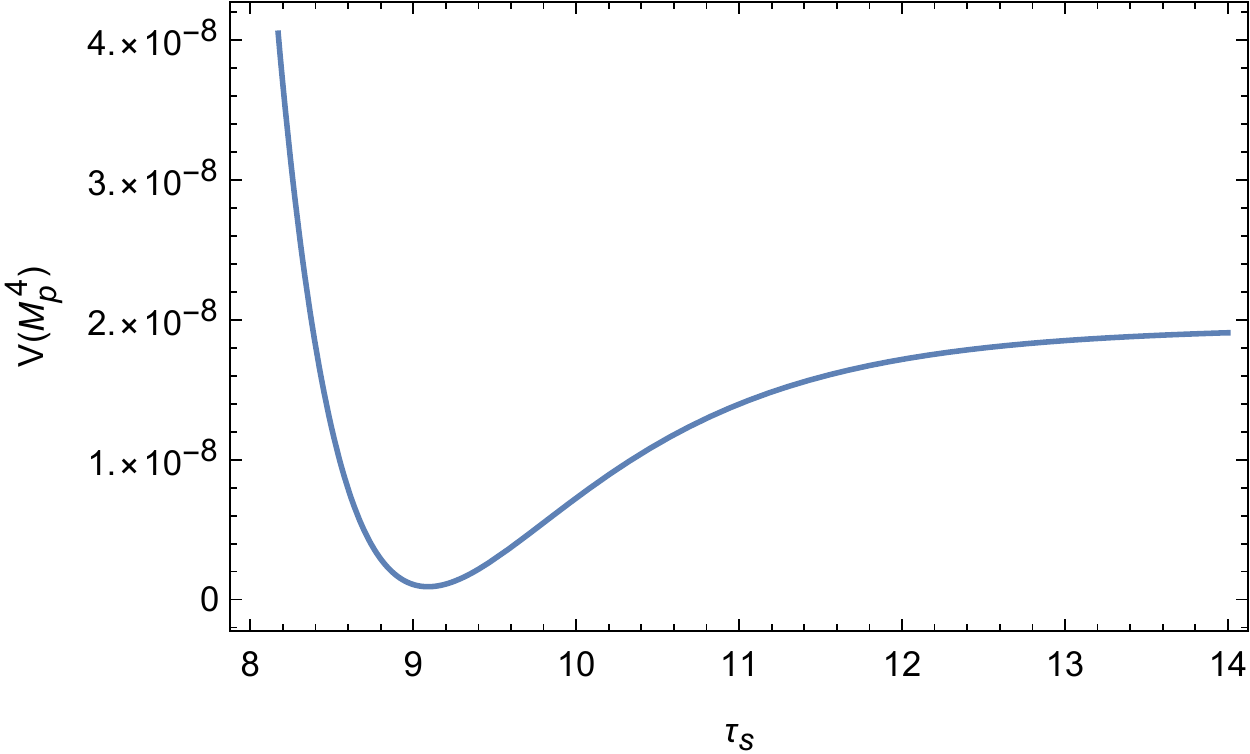}
\end{subfigure}
\caption{Plots of the potential (\ref{eq:LVS_full_potential}) in each of the 3 directions $(\zeta,\tau_b,\tau_s)$, for the parameter set in Table \ref{Tab2}, around the minimum (left) and the saddle point (right).}
 \label{fig6_E1}
\end{figure}

\begin{figure}[]
  	\centering
  	\includegraphics[width=0.7\linewidth]{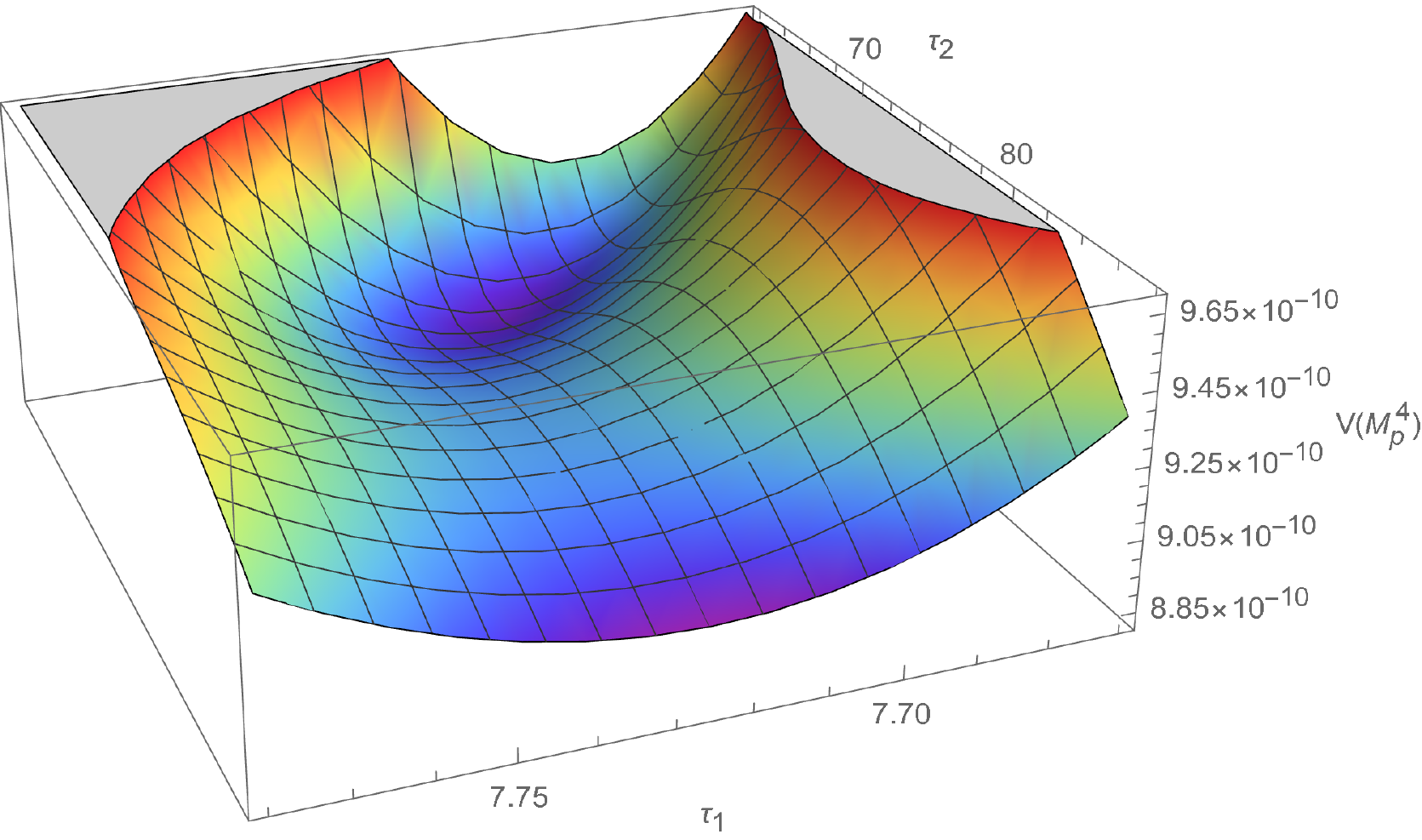}
	\caption{Plot of the potential (\ref{eq:LVS_full_potential}) in the $(\tau_1,\tau_2)$ plane (which is just a rotation of the $(\tau_b,\tau_s)$ plane aligned with the eigenvectors of the Hessian matrix), for the parameters in Table \ref{Tab2}.}
 \label{fig:Example_plot3d}
\end{figure}

In Figs.\ref{fig6_E1}-\ref{fig:Example_plot3d} we show the plots of the potential in each direction and in the plane $(\tau_s,\tau_b)$. 


%% file: Sections/conclusions.tex
\newpage
\section{Conclusions}
\label{S:Conclusions}

We have revisited moduli stabilisation for type IIB Calabi-Yau flux compactifications that include an $\antiD$-brane at the tip of a KS throat, a warped deformed conifold.  In particular, we have focussed on the stabilisation of the conifold's deformation modulus, including its coupling with the Calabi-Yau bulk volume modulus. It is important to study the coupled conifold-volume-moduli system because the flux-induced mass of the conifold modulus is extremely light, and it may be destabilised by the ingredients introduced to stabilise the volume modulus.

Indeed, it has recently been observed that the $\antiD$-brane uplifting potential energy density can destabilise the conifold modulus, which would otherwise be stabilised by fluxes \cite{upliftingrunaways2019} (see also \cite{Blumenhagen:2019qcg, Bena:2019sxm, Dudas:2019pls, lisa2019, Demirtas:2020ffz}).  This happens because the $\antiD$-brane energy density is warped down, and the background warp factor at the tip depends on the vev of the conifold modulus. The full scalar potential descends from fluxes and quantum corrections, as well as the uplift term,  and can be computed within the low energy effective field theory from an appropriate K\"ahler potential and superpotential.  A key object in the discussion is the K\"ahler metric for the conifold modulus, which was worked out in \cite{douglas2007warping} and presented in detail in Appendix \ref{sec:complexstructuremoduli},
\begin{align}
    G_{z\Bar{z}}
    = \frac{1}{\pi||\Omega||^2} 
    \left(\log\frac{\Lambda_{0}^3}{|z|}
    + \frac{c' }{(2\pi)^4}\frac{1}{\V^{2/3}} \frac{(g_sM)^2}{|z|^{4/3}}
    \right), \nonumber
\end{align}
where the $c'$ term originates from the warping and induces a mixing between the volume and conifold modulus.  In all previous works, this warping term in the metric has been assumed to dominate.  In this work, we  considered a new, previously unexplored, region of parameter space, where the warping term is subdominant in the metric.

After reviewing the computation of the conifold-volume modulus potential, we began in Section \ref{DeformStab_sec} by studying the stabilisation of the conifold modulus by the flux superpotential, deferring the full stabilisation including the volume modulus to Section \ref{LVS_section}.  In the case of strong warping, we confirmed the well-known approximate GKP solution with $|z| \sim e^{-\frac{2\pi K}{g_s M}}$, present when the bound $\sqrt{g_s}M \gtrsim 6.8$ is satisfied \cite{upliftingrunaways2019}.  Then we considered the weakly-but-still-warped case.  Here, we found a qualitatively new solution, which takes the form $|z|^{4/3} \sim\V^{-2/3}$, and which exists only when the anti-brane is present and when the new bound $\sqrt{g_s}M \gtrsim 13.6$ is satisfied.  
To stay in the weakly-warped regime, large volumes and long throats are preferred, $\V^{2/3} \, \Lambda_0^4 \gg e^{\frac{8\pi K}{3 g_s M}}$.

Having identified a new region of parameter space, with large volumes, we proceeded in Section \ref{LVS_section} to embed the above conifold modulus stabilisation into the Large Volume Scenario to achieve a stabilised volume modulus.  To this end, it suffices to extend the K\"ahler modulus sector to the large and small moduli of a ``Swiss cheese'' Calabi-Yau and add the leading $\alpha'$ correction to $\mathcal{K}$ and a non-perturbative superpotential, the interplay of which can lead to K\"ahler moduli stabilisation.  After working out the general scalar potential for the three complex-moduli system, we again analyse in turn the warping dominated regime and the regime of subdominant warping.  For the strongly warped regime, we recover the results of \cite{LVSdS:2010.15903}, and moreover show that the final dS solution is consistent with the said regime.  For the weakly-but-still-warped regime, we find new metastable dS solutions with all moduli stabilised, consistent with the weak coupling, supergravity approximation and the subdominance of the warping. This new solution only exists when the $\antiD$-brane is present.  Note that in our new de Sitter solution, the brane potential is still exponentially suppressed, but this now originates not from GKP but rather from the exponentially large volume of the Large Volume Scenario, with $|z|^{4/3} \sim \V^{-2/3}$. The hierarchy of scales between the UV and IR is always very small for this solution.

The final vevs and mass hierarchies are not very different in the two regimes, being distinguished in our examples mainly by the vev of the conifold modulus, $|z|$, which is larger in the new regime, and the vev of the volume modulus, $c$, which ends up somewhat smaller.  In the end, the tadpole numbers from the throat fluxes are $MK \sim {\mathcal{O}}(100)$, similar to the LVS solution \cite{LVSdS:2010.15903}, and there also remains the issue of tadpole contributions from bulk fluxes \cite{Betzler2019,tadpoleProblem,Ishiguro2021}.  In fact, the parameter space where we have found the new dS solutions have somewhat large values $\mathcal{O}(10^2 - 10^3)$ for $W_0$, from the heavy bulk moduli, and for $A$, from the leading non-perturbative term to stabilise the Swiss Cheese moduli.   The coefficient $W_0$ is usually taken to come from bulk fluxes, although other sources such as from non-perturbative effects in heavy K\"ahler moduli are also possible\footnote{See \cite{Demirtas2019sip,Demirtas:2020ffz,Blumenhagen_smallflux,Honma2021klo} for progress in achieving the small values of $W_0$ necessary for KKLT.}.  The coefficient $A$ is related to a one-loop instanton determinant in the case of E3 branes or to the cutoff of the effective gauge theory, possibly enhanced by threshold effects, for gaugino condensation on wrapped D7-branes (see \cite{Denef:2005mm} for F-theory examples with large $A$). Other possible control issues that have been raised recently for KKLT in \cite{Carta_2019, Gao:2020xqh, Carta:2021lqg}, that is the throat not fitting into the bulk and the appearance of pathological bulk singularities, are not a danger here.  

An interesting open issue is a ten-dimensional understanding of our new de Sitter solutions\footnote{See \cite{Moritz2017xto,Hamada2019ack,Gautason2019jwq,Cribiori2019hod,Bena2019mte,Kachru2019dvo} for progress towards a ten-dimensional understanding of KKLT.}.  When studying the conifold deformation modulus, we start from the non-compact ten-dimensional Klebanov-Strassler solution cut off at some radial distance $r_{UV}$ and embeded into a GKP Calabi-Yau flux compactification.  As has been emphasised in \cite{Dudas:2019pls}, the deformation parameter is not a flat direction in the KS/GKP solution and is fixed to its supersymmetric value, $|z| \sim \Lambda_{0}^3 e^{-\frac{2\pi K}{g_s M}}$, which can also be understood from our consistency condition, (\ref{eq:relation_K_Lambda0}), coming from the B-cycle NSNS flux quantisation along the cut off throat.  Therefore, one might not trust the effective field theory far from this value for $|z|$, although doing so does lead to results that are consistent with the backreacted analysis in \cite{Bena:2019sxm} and also \cite{lisa2019}.  In our new solution, $|z|$ is about two orders of magnitude smaller than the corresponding GKP value.  At the same time, we have introduced several ingredients that go beyond KS/GKP into our effective field theory -- quantum corrections, localised sources, and four-dimensional effects such as gaugino-condensation -- which we have not fully described in ten-dimensions.  Indeed, the new dS solution would not exist without the $\antiD$-brane at the throat tip and we could only study the influence of the $\antiD$-brane on the conifold deformation parameter using the four-dimensional effective field theory.  
Despite the relatively large flux numbers/D3-branes at the tip, the induced mass for $|z|$ is so light that a single $\antiD$-brane and a large bulk volume can influence significantly its dynamics and vev.  Our final de Sitter solution has an ameliorated tadpole problem, no bulk-singularity problem, and is consistent with the Kaluza-Klein truncation, $\alpha'$ and string-loop expansions and supergravity description.  However, without a ten-dimensional description, we cannot be certain that it lies in the string theory landscape and is not further evidence of a swampland conspiracy.

%% file: Sections/cs_moduli_metric.tex
\section{Complex structure moduli metric}
\label{sec:complexstructuremoduli}

Here we review the computation of $G_{S\bar{S}}$ for the conifold deformation modulus, $S$, using the KS metric, making explicit the appearance of the volume modulus. We follow closely \cite{douglas2007warping}. 

The metric in the complex structure moduli space can be be computed using 

\begin{align}
    G_{\alpha\Bar{\beta}} = \frac{i\int h~\chi_\alpha\wedge\chi_{\Bar{\beta}}}{i\int h~\Omega\wedge\Bar{\Omega}},
    \label{eq:CS_moduli_metric}
\end{align}

\noindent where now, $h = 1 + \frac{e^{-4A_0(y)}}{c(x)}$. This corresponds to the Kahler potential
\begin{align}
    \mathcal{K}_{cs} = -\log\left(\frac{i}{\kappa_4^6}\int h~\Omega\wedge\Bar{\Omega}\right).
\end{align}

In our setup, it is assumed that all complex structure moduli are stabilised in the UV, i.e. in the bulk, except for the deformation modulus $S$ that governs the Klebanov-Strassler geometry and lives in the highly-warped region. In particular, this means we can split the Kahler potential into two different contributions
\begin{align}
    \mathcal{K}_{cs} &= -\log\left(\frac{i}{\kappa_4^6}\int h~\Omega\wedge\Bar{\Omega} \right) \nn\\
    &= -\log\left(\frac{i}{\kappa_4^6}\int_{bulk} h~\Omega\wedge\Bar{\Omega} + \frac{i}{\kappa_4^6}\int_{conifold} h~\Omega\wedge\Bar{\Omega} \right) \nn\\
    &\approx -\log\left(\frac{i}{\kappa_4^6}\int_{bulk} \Omega\wedge\Bar{\Omega} + \frac{i}{\kappa_4^6}\int_{conifold} h~\Omega\wedge\Bar{\Omega} \right) \nn\\
    &= -\log\left(\frac{i}{\kappa_4^6}\int_{bulk} \Omega\wedge\Bar{\Omega}
    \left(1 + \frac{i\int_{conifold} h~\Omega\wedge\Bar{\Omega}}{i\int_{bulk} \Omega\wedge\Bar{\Omega}} \right)\right) \nn\\
    &= -\log\left(\frac{i}{\kappa_4^6}\int_{bulk} \Omega\wedge\Bar{\Omega}\right)
    - \log\left(1 + \frac{i\int_{conifold} h~\Omega\wedge\Bar{\Omega}}{i\int_{bulk} \Omega\wedge\Bar{\Omega}} \right) \nn\\
    &\approx \mathcal{K}_{cs}^{UV} +  \frac{e^{\mathcal{K}_{cs}^{UV}}}{\kappa_4^6} i\int_{conifold} h~\Omega\wedge\Bar{\Omega}\nn \\
    &= \mathcal{K}_{cs}^{UV} + \mathcal{K}(S,\Bar{S})\,,
\end{align}

\noindent where the first approximation follows from $h\approx 1$ in the bulk and the second assumes the contribution from the bulk is much bigger than the one from the conifold. 
\begin{align}
    \mathcal{K}_{cs}^{UV} &= -\log\left(\frac{i}{\kappa_4^6}\int_{bulk} \Omega\wedge\Bar{\Omega}\right) 
    = -\log\left(\frac{||\Omega||^2V_{6}}{\kappa_4^6}\right)\,.
\end{align}

We can now compute the conifold contribution to the metric following the computations in \cite{douglas2007warping}
\begin{align}
    G_{S\Bar{S}} = \frac{e^{\mathcal{K}_{cs}^{UV}}}{\kappa_4^6}i\int_{conifold} h~\chi_S\wedge\chi_{\Bar{S}}.
\end{align}
For the KS metric, the (2,1)-form $\chi_S$ is given by
\begin{align}
    \chi_S = g^3\wedge g^4\wedge g^5 + d[F(\eta)(g^1\wedge g^3 + g^2\wedge g^4 )] - id[f(\eta)(g^1\wedge g^2) + k(\tau)(g^3\wedge g^4)],
\end{align}
where the functions $f,k,F$ were computed in \cite{KS2000supergravity} 
\begin{align}
    F(\eta) = \frac{\sinh{\eta}-\eta}{2\sinh{\eta}} ,
    && f(\eta) = \frac{\eta\coth{\eta}-1}{2\sinh{\eta}}(\cosh{\eta}-1) ,
    && k(\eta) = \frac{\eta\coth{\eta}-1}{2\sinh{\eta}}(\cosh{\eta}+1).
    \label{eq:ansatz_functions_KS}
\end{align}
It will be useful to look at the limits $\eta\rightarrow 0$ and $\eta\rightarrow\infty$ of these functions:
\begin{align}
    &F(\eta\rightarrow 0) = \frac{\eta^2}{12}\,,
    && f(\eta\rightarrow 0) = \frac{\eta^3}{12}\,,
    && k(\eta\rightarrow 0) = \frac{\eta}{3}\,, \\
    &F(\eta\rightarrow\infty) = \frac{1}{2} -\eta e^{-\eta}\,,
    && f(\eta\rightarrow\infty) = \frac{\eta}{2}\,,
    && k(\eta\rightarrow\infty) = \frac{\eta}{2}\,,
\end{align}
Surprisingly the combination we need, $\chi_S\wedge\chi_{\Bar{S}}$, is a total $\eta$-derivative
\begin{align}
    \chi_S\wedge\chi_{\Bar{S}} = -\frac{2i}{64\pi^4}d\eta \wedge \left(\prod_i g^i\right) \frac{d}{d\eta}[f+F(k-f)],
\end{align}
from which we find for $G_{S\Bar{S}}$ (with all the integrals over the conifold region)
\begin{align}
    G_{S\Bar{S}} &= \frac{e^{\mathcal{K}_{cs}^{UV}}}{\kappa_4^6}i\int h~\chi_S\wedge\chi_{\Bar{S}} \nn\\
    &= \frac{i}{||\Omega||^2V_{6}} \int h~\chi_S\wedge\chi_{\Bar{S}} \nn\\
    &= \frac{1}{||\Omega||^2V_{6}} \frac{2}{64\pi^4} \int h~ d\eta \wedge \left(\prod_i g^i\right) \frac{d}{d\eta}[f+F(k-f)] \nn\\
    &= \frac{2}{64\pi^4||\Omega||^2V_{6}} 
    \left(\int \prod_i g^i\right)
    \int d\eta ~h~ \frac{d}{d\eta}[f+F(k-f)] \nn\\
    &= \frac{2}{\pi||\Omega||^2V_{6}}
    \left(\int d\eta \frac{d}{d\eta}\{h~[f+F(k-f)]\}
    - \int d\eta \frac{dh}{d\eta}[f+F(k-f)]
    \right) ,
\end{align}

\noindent where we preformed the $\eta$ integral by parts and used $\int \prod_i g^i = 64\pi^3$. 

We are left with an integral in $\eta$, which corresponds to the radial coordinate of the conifold. We introduce a cutoff scale where the conifold connects to the bulk. For the coordinate defined in the limit $\eta\rightarrow\infty$,
\begin{align}
    r_\infty^2 = \frac{3}{2^{5/3}}s^{2/3}e^{2\eta/3},
\end{align}
the cutoff scale $\Lambda_{UV}$ is defined as
\begin{align}
    \Lambda_{UV}^2 \equiv r_{UV}^2 =  \frac{3}{2^{5/3}}s^{2/3}e^{2\eta_{\Lambda}/3} 
    \implies \eta_\Lambda = \frac{3}{2}\log\frac{2^{5/3}}{3} + \log\frac{\Lambda_{UV}^3}{s}\,,
    \label{eq:Lambda0_def}
\end{align}

\noindent where $\eta_\Lambda$ is the maximum value of $\eta$, i.e. the upper bound of our integral where the conifold connects with the bulk. We see that the first term in the integral is just a boundary term, so it suffices to evaluate $h~[f+F(k-f)]$ at $\eta\rightarrow 0$ and $\eta\rightarrow\eta_\Lambda$ (where we can think of $\eta_\Lambda \gg 1$ and use the approximations for $\eta\rightarrow\infty$). It is useful to recall the warp factor for the deformed conifold, now written in terms of the complex structure $s=|S|=\epsilon^2$
\begin{align}
    e^{-4A_0(y)} = 2^{2/3}\frac{(\alpha'g_sM)^2}{s^{4/3}}I(\eta),
    &&
    I(\eta)\equiv \int_{\eta}^{\infty} dx ~ \frac{x\coth(x) - 1}{\sinh^2{x}}(\sinh(2x)-2x)^{1/3}\,,
\end{align}

\noindent At $\eta\rightarrow 0$, $I(0)\approx 0.718$ and $f+F(k-f)=0$, and at $\eta\rightarrow\eta_\Lambda$, in the bulk, $h\approx 1$ and
\begin{align}
    f+F(k-f) \approx \frac{\eta_\Lambda}{2} =  \frac{3}{4}\log\frac{2^{5/3}}{3} + \frac{1}{2}\log\frac{\Lambda_{UV}^3}{s}.
\end{align}
As for the second term, all we need is the derivative of $h$ 

\begin{align}
    \frac{dh}{d\eta}
    =\frac{1}{c(x)}\frac{de^{-4A_0(y)}}{d\eta} 
    &= 2^{2/3}\frac{(\alpha'g_sM)^2}{s^{4/3}} \frac{1}{c(x)}\frac{dI(\eta)}{d\eta}\nn\\
    &= -4\times 2^{2/3}\frac{(\alpha'g_sM)^2}{s^{4/3}} \frac{1}{c(x)}\frac{f+F(k-f)}{(\sinh(2\eta)-2\eta)^{2/3}}\,,
\end{align}

\noindent and so the second integral (which is well approximated by taking $\eta_\Lambda\to\infty$) gives

\begin{align}
    -4\times  2^{2/3}\frac{(\alpha'g_sM)^2}{s^{4/3}} \frac{1}{c(x)}\int_0^{\eta_\Lambda} \frac{[f+F(f-k)]^2}{(\sinh(2\eta)-2\eta)^{2/3}} 
    \approx 0.093\times  (-4\times 2^{2/3})\frac{(\alpha'g_sM)^2}{s^{4/3}} \frac{1}{c(x)}.
    \label{eq:integral_metric_GSS}
\end{align}

Hence, the metric becomes

\begin{align}
    G_{S\Bar{S}} 
    &= \frac{1}{\pi||\Omega||^2V_{6}} 
    \left(\frac{3}{2}\log\frac{2^{5/3}}{3} + \log\frac{\Lambda_{UV}^3}{s}
    + 0.093\times 8 \times 2^{2/3}\times \frac{(\alpha'g_sM)^2}{s^{4/3}}\frac{1}{c(x)}
    \right) \,.
\end{align}
Defining the constant $c' \approx 0.093\times 8 \times 2^{2/3} \approx 1.18$ and for $s\ll\Lambda_{UV}^3$ (which is equivalent to the assumption $\eta_\Lambda\gg 1$), we can neglect the first term 
\begin{align}
    G_{S\Bar{S}}
    = \frac{1}{\pi||\Omega||^2V_{6}} 
    \left(\log\frac{\Lambda_{UV}^3}{s}
    + c' \frac{(\alpha'g_sM)^2}{s^{4/3}}\frac{1}{c(x)}
    \right).
    \label{eq:appendixC_GSS}
\end{align}
%
This metric corresponds to the Kahler potential
\begin{align}
   \mathcal{K}(S,\Bar{S}) =\frac{1}{\pi||\Omega||^2V_{6}} \left(|S|^2\left(\log\frac{\Lambda_{UV}^3}{|S|} + 1\right) + \frac{9c'(\alpha'g_sM)^2}{c(x)}|S|^{2/3}\right).
\end{align}

We can now make a field redefinition, introducing a dimensionless deformation modulus $z=S/l_s^3$ and identifying the volume modulus $c(x)=\V^{2/3}$
\begin{align}
    \mathcal{K}(z,\Bar{z}) =\frac{l_s^6
    }{\pi||\Omega||^2V_{6}}\left(|z|^2\left(\log\frac{\Lambda_0^3}{|z|} + 1\right) + \frac{9c'(g_sM)^2}{(2\pi)^4\V^{2/3}}|z|^{2/3}\right),
\end{align}
where now $\Lambda_0=\Lambda_{UV}/l_s$ is expressed in string units of $l_s$. With this redefinition we can say that ``$z$ is small'', i.e. the dimensionless quantity $|S|/l_s^3\ll 1$ or $|S|$ is small in string units.

Finally, we can make the choice $V_{6}=l_s^6$. Keeping it allows us to keep track of $V_{6}$ and remember where this factor comes from but, being just a volume integral, we can always choose to normalize it in such a way, since the volume modulus keeps the overall volume dependence. 